\documentclass[10pt]{article}
\usepackage[latin1]{inputenc}
\usepackage{amsmath}
\usepackage{amsfonts}
\usepackage{amssymb}
\usepackage{mathrsfs}

\usepackage{MnSymbol}

\usepackage[cal=boondox,scr=boondoxo]{mathalfa}

\usepackage{float}
\usepackage{subfig}
\usepackage{fancybox,graphicx}
\usepackage{subfig}
\usepackage{caption}
\usepackage{color}
\usepackage{authblk}

\usepackage[colorlinks]{hyperref}
\usepackage{hyperref}

\newcommand{\doi}[1]{{doi:~\href{https://doi.org/#1}{\nolinkurl{#1}}}\rmFullStop}

\newcommand*{\rmFullStop}{\rmifnextchar{.}{}{}}

\makeatletter
\newcommand{\rmifnextchar}[3]{%
  \begingroup
  \ltx@LocToksA{\endgroup#2}%
  \ltx@LocToksB{\endgroup#3}%
  \ltx@ifnextchar{#1}{%
    \def\next{\the\ltx@LocToksA}%
    \afterassignment\next
    \let\scratch= %
  }{%
    \the\ltx@LocToksB
  }%
}
\makeatother %doi command

\usepackage{accents}
\usepackage[titletoc,title]{appendix}
\usepackage{cite}

\usepackage{mathtools}

\usepackage[top=2in, bottom=1.5in, left=1in, right=1in]{geometry}

\title{Electric Vector Potential Approach in Electrostatics : The Surface Electrode} 
 
\author[1]{Robert Salazar}
\author[2,3]{Camilo Bayona-Roa}
\author[1]{Gabriel T\'ellez}
\affil[1]{Departamento de F\'isica, Universidad de los Andes - Bogot\'a, Colombia}
\affil[2]{Departamento de Ciencias B\'asicas, Universidad ECCI - Bogot\'a, Colombia}
\affil[3]{Centro de Ingenier\'ia Avanzada Investigaci\'on y Desarrollo, CIAID- Bogot\'a, Colombia}
\begin{document}

    \maketitle

\begin{abstract}
Electric vector potential $\Theta(\boldsymbol{r})$ is a legitimate but rarely used tool to calculate the steady electric field in free-charge regions. Commonly, it is preferred to employ the scalar electric potential $\Phi(\boldsymbol{r})$ rather than $\Theta(\boldsymbol{r})$ in most of the electrostatic problems. However, the electric vector potential formulation can be a viable representation to study certain systems. One of them is the surface electrode SE, a planar finite region $\mathcal{A}_{-}$ kept at a fixed electric potential with the rest grounded including a gap of thickness $\nu$ between electrodes. In this document we use the \textit{Helmholtz Decomposition Theorem} and the electric vector potential formulation to provide integral expressions for the surface charge density and the electric field of the SE of arbitrary contour $\partial\mathcal{A}$. We also present an alternative derivation of the result found in [M. Oliveira and J. A. Miranda 2001 Eur. J. Phys. 22 31] for the gapless ($\nu=0$) surface electrode GSE without invoking any analogy between the GSE and magnetostatics. It is shown that electric vector potential and the electric field of the gapped circular SE at any point can be obtained from an average of the gapless solution on the gap.
\end{abstract}

\section{Introduction}
\begin{figure}[H]
\centering %system.pdf
\includegraphics[width=0.6\textwidth]{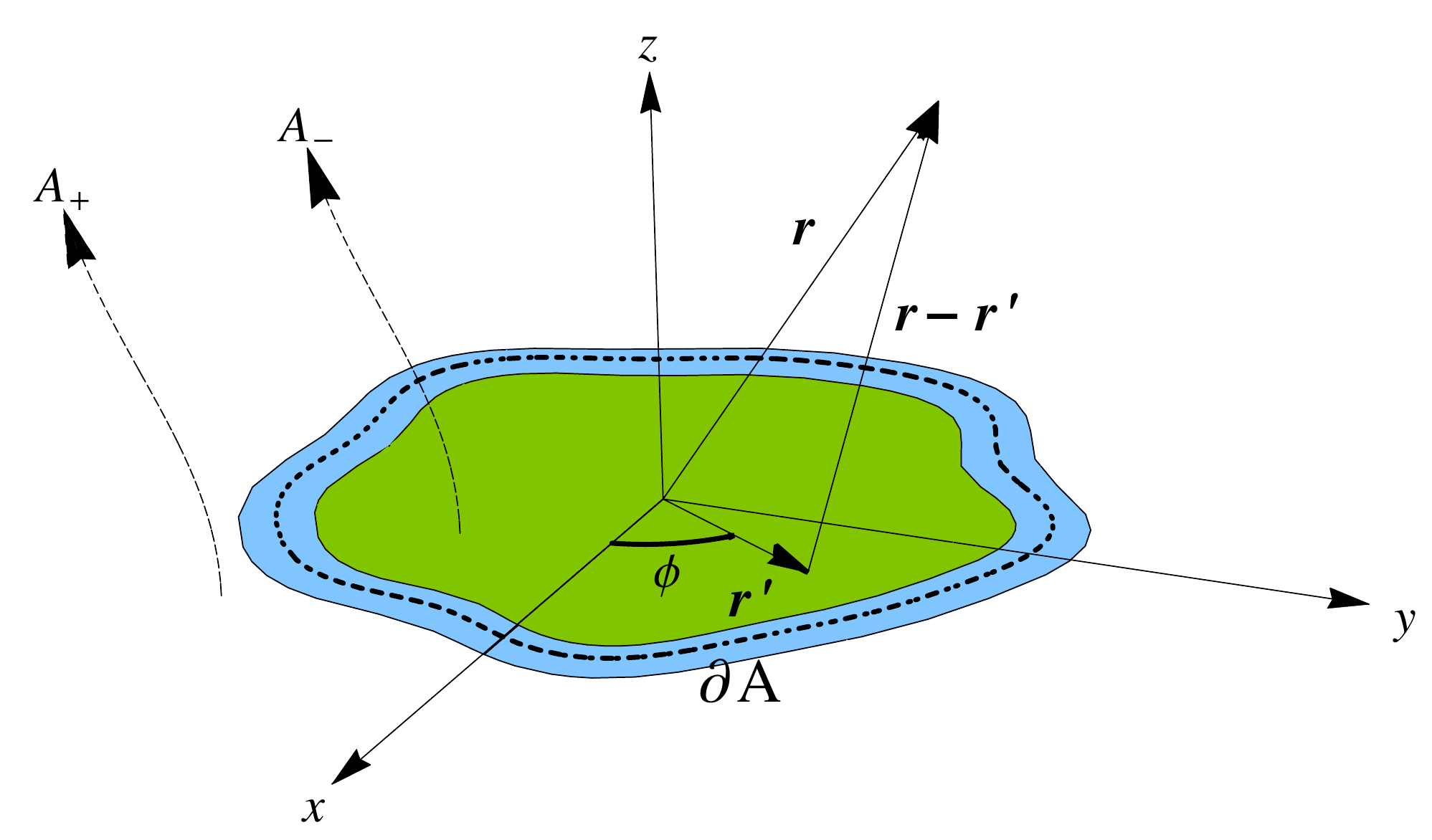}
    \caption[The system.]{The SE is composed by two flat sheets at different electric potential including a gap represented by the gray region in the neighbourhood of $\partial \mathcal{A}$.}
\label{theSystemFig}
\end{figure}

Calculation of electric field via the scalar electric potential $\Phi(\boldsymbol{r})$ is an standard procedure in in electrostatics. However, the steady electric field in free charge regions is an irrotational field $\nabla \times \boldsymbol{E} = 0$ but also divergence free $\nabla \cdot \boldsymbol{E} = 0$. This enables to associate an electric vector potential to the electric field
\[
\boldsymbol{E} = \nabla\times\Theta(\boldsymbol{r})  \hspace{0.5cm}\textbf{if}\hspace{0.5cm}\rho(\boldsymbol{r}) = 0
\]
where $\rho(\boldsymbol{r})$ is the charge density. This implies that vector electric potential satisfies the Laplace equation   
\[
\nabla^2\Theta(\boldsymbol{r}) = 0    \hspace{0.5cm}\textbf{if}\hspace{0.5cm}\rho(\boldsymbol{r}) = 0
\]
when a Coulomb-like gauge condition $\nabla \cdot \Theta(\boldsymbol{r}) = 0$ is imposed. Even when electric vector potential is a valid choice to place some electrostatics problems, it is commonly preferred the scalar potential. A similar situation occurs in magnetostatics where vector magnetic potential is usual a better choice than the scalar magnetic potential in free-current regions. However, there are systems where expressing $\boldsymbol{E}$ in terms of the electric vector potential is not only an available choice, but also a advantageous one. This is the case of the gapped Surface Electrode SE, a conductor infinite sheet on the $xy$-plane having a region $\mathcal{A}_{-}$ at constant potential $V_o$, a sheet grounded $\mathcal{A}_{+}$ and a gap in between (see Fig.~\ref{theSystemFig}).      

In this document, the gap is defined as follows
\[
\mathcal{G} = \{ (r,\phi,0) : \hspace{0.25cm} \mathscr{R}(\phi) - \nu/2 < r < \mathscr{R}(\phi) + \nu/2 \hspace{0.25cm} \forall \hspace{0.25cm} \phi \in [0,2\pi) \}
\]
with $(\mathscr{R}(\phi),\phi)$ and $(\mathscr{R}(\phi) \pm \nu/2,\phi)$ the parametric representations of contours $\partial A$ and $\partial A_{\pm}$ respectively. The problem consists in solving the Laplace's equation $\nabla^2 \Phi(\boldsymbol{r}) = 0$, $\boldsymbol{r} \in \mathfrak{D}=\left\{\boldsymbol{r} \in \mathbb{R}^3 : z > 0 \right\}$ subjected to the boundary conditions
\begin{equation}
\Phi(\boldsymbol{r}) = V_o  \hspace{0.5cm} \mbox{\textbf{if}} \hspace{0.5cm} \boldsymbol{r} \in \mathcal{A}_{-} \subset \left\{\boldsymbol{r} \in \mathbb{R}^2 : z = 0 \right\}, \hspace{0.25cm}
\Phi(\boldsymbol{r}) = 0 \hspace{0.5cm} \mbox{\textbf{if}} \hspace{0.5cm} \boldsymbol{r} \in \left\{\boldsymbol{r} \in \mathbb{R}^2 : z = 0 \right\}\setminus\mathcal{A}_{-}\cup\mathcal{G},
\label{DirichletBoundaryConditionsEq}
\end{equation}
and 
\begin{equation}
\lim_{z \rightarrow 0} \frac{\partial \Phi(\boldsymbol{r})}{\partial z} = 0 \hspace{0.5cm} \mbox{\textbf{if}} \hspace{0.5cm} \boldsymbol{r} \in \mathcal{G}
\label{NeumannBoundaryConditionsEq}
\end{equation}
in order to determine the electric in the $\mathbb{R}^3$ space, and the surface charge density distribution on the $xy$-plane. This system plays an important role in the study of Surface-electrode (SE) Radio Frequency ion traps which can be modelled as an infinite plane gaplessly covered by an array of SE. Those SE ion traps are a promising candidates to build ion-trap networks suitable for large-scale quantum processing \cite{chiaverini2005surface,leibfried2005surface,seidelin2006microfabricated,daniilidis2011fabrication,kim2011surface,hong2017experimental,mokhberi2017optimised,tao2018fabrication}. There are several works describe analytic treatments including the SE in diverse situations : rectangular strip electrode held at constant \cite{house2008analytic}, Ring-shaped SE traps  \cite{wesenberg2008electrostatics, schmied2010electrostatics}, and the gapless SE with angular dependent potential \cite{salazar2019AngularDependentSE}.

In this document the Hemholtz Descomposition Theorem in combination with the Green's theorem are used to find the vector electric potential of the SE, recovering the Biot-Savart like (BSL) law for the electric field found in Ref.~\cite{oliveira2001biot} without employing any analogy with magnetostatics. The vector electric potential will be also used to find the charge density of the circular SE and showing that it is a globally neutral system.

\section{Helmholtz decomposition theorem}
\label{HelmholtzDecompositionTheoremSectionLabel}
The Helmholtz theorem \cite{von33uber}\cite[p.~147]{sommerfeld1950mechanics}\cite{cong1993helmholtz,gui2007rigorous} states that a twice continuously differentiable field $\boldsymbol{P}(\boldsymbol{r})$ on a bounded region $\Gamma \subset \mathbb{R}^3$ can be decomposed into a irrotational (curl-free) plus a solenoidal (divergence-free) fields as follows

\[
\boldsymbol{P}(\boldsymbol{r}) = - \nabla \zeta(\boldsymbol{r}) + \nabla \times \boldsymbol{\Omega}(\boldsymbol{r}) 
\]
where
\[
\zeta(\boldsymbol{r}) = \frac{1}{4\pi} \int_{\Gamma} \frac{\nabla'\cdot\boldsymbol{P}(\boldsymbol{r}')}{|\boldsymbol{r}-\boldsymbol{r}'|} d^3\boldsymbol{r} - \frac{1}{4\pi} \int_{\partial \Gamma} \frac{\hat{n}'\cdot\boldsymbol{P}(\boldsymbol{r}')}{|\boldsymbol{r}-\boldsymbol{r}'|} d^2\boldsymbol{r}
\]
and 
\[
\boldsymbol{\Omega}(\boldsymbol{r}) = \frac{1}{4\pi} \int_{\Gamma} \frac{\nabla'\times\boldsymbol{P}(\boldsymbol{r}')}{|\boldsymbol{r}-\boldsymbol{r}'|} d^3\boldsymbol{r} - \frac{1}{4\pi} \int_{\partial \Gamma} \frac{\hat{n}'\times\boldsymbol{P}(\boldsymbol{r}')}{|\boldsymbol{r}-\boldsymbol{r}'|} d^2\boldsymbol{r}
\]
when $\boldsymbol{P}(\boldsymbol{r})$ decreases asymptotically ($r\rightarrow\infty$). Decomposing the electric field $\boldsymbol{P}(\boldsymbol{r}) = \boldsymbol{E}(\boldsymbol{r})$ in the region $\mathfrak{D}$ then
\begin{equation}
\boldsymbol{E}(\boldsymbol{r}) = - \nabla \zeta(\boldsymbol{r}) + \nabla \times \boldsymbol{\Omega}(\boldsymbol{r})
\label{hemholtzDescompositionEFieldEq}    
\end{equation}
where 
\[
\zeta(\boldsymbol{r}) = \frac{1}{4\pi \epsilon_o} \int_{\Gamma} \frac{\rho(\boldsymbol{r}')}{|\boldsymbol{r}-\boldsymbol{r}'|} d^3\boldsymbol{r} - \frac{1}{4\pi} \int_{\partial \Gamma} \frac{\hat{n}'\cdot\boldsymbol{E}(\boldsymbol{r}')}{|\boldsymbol{r}-\boldsymbol{r}'|} d^2\boldsymbol{r}
\]
where we have used the Gauss's law $\nabla\cdot\boldsymbol{E}(\boldsymbol{r})=\rho(\boldsymbol{r})/\epsilon_o$. Here $\hat{n}'(\boldsymbol{r}')$ is the normal outward vector of $\partial\mathcal{D}$, and it is $\hat{n}'=-\hat{z}$ on $\partial\mathcal{D}_{z=0}$. Note that volume integral is zero since there is no charge density in $\mathcal{D}$. However, there is an unknown but not null charge density spread on $\partial \mathcal{D}_{z=0}$ given by 
\begin{equation}
\sigma(x,y) = 2 \epsilon_o \lim_{z \to 0^{+}} E_z(x,y,z)
\label{surfaceChargeDensityLimitEq}
\end{equation}
therefore
\[
\zeta(\boldsymbol{r}) =  \frac{1}{8\pi\epsilon_o} \int_{\mathbb{R}^2} \frac{\sigma(x',y')}{|\boldsymbol{r}-\boldsymbol{r}'|} dx'dy'.
\]
On the other hand, the field $\boldsymbol{\Omega}(\boldsymbol{r})$ can be obtained from
\[
\boldsymbol{\Omega}(\boldsymbol{r}) = \frac{1}{4\pi} \int_{\mathfrak{D}} \frac{\nabla'\times\boldsymbol{E}(\boldsymbol{r}')}{|\boldsymbol{r}-\boldsymbol{r}'|} d^3\boldsymbol{r} - \frac{1}{4\pi} \int_{\partial \mathfrak{D}} \frac{\hat{n}'\times\boldsymbol{E}(\boldsymbol{r}')}{|\boldsymbol{r}-\boldsymbol{r}'|} d^2\boldsymbol{r}.
\]
which can be simplified as follows 
\begin{equation}
\boldsymbol{\Omega}(\boldsymbol{r}) = \frac{1}{4\pi} \int_{\mathbb{R}^2} \frac{\hat{z}\times\boldsymbol{E}(\boldsymbol{r}')}{|\boldsymbol{r}-\boldsymbol{r}'|} dx'dy'.
    \label{OmegaUsefulEq}
\end{equation}
because the steady electric field is irrotational. Since $\hat{z}\times\boldsymbol{E}(\boldsymbol{r}') = (\partial_y \Phi(\boldsymbol{r}), -\partial_x \Phi(\boldsymbol{r}), 0)$ in Cartesian coordinates, then the components of $\boldsymbol{\Omega}(\boldsymbol{r}) = (\Omega_x(\boldsymbol{r}),\Omega_x(\boldsymbol{r}),0)$ are given by  
\[
\Omega_x(\boldsymbol{r}) = \frac{1}{4\pi} \int_{\mathbb{R}^2} \frac{1}{|\boldsymbol{r}-\boldsymbol{r}'|} \partial_y \Phi(x',y',0) dx'dy' \hspace{0.5cm} \mbox{and} \hspace{0.5cm} \Omega_y(\boldsymbol{r}) = -\frac{1}{4\pi} \int_{\mathbb{R}^2} \frac{1}{|\boldsymbol{r}-\boldsymbol{r}'|} \partial_x \Phi(x',y',0) dx'dy'.
\]
The $x$-component of $\boldsymbol{\Omega}(\boldsymbol{r})$ can be simplified by performing a partial integration with respect the $y'$-coordinate 
\[
\Omega_x(\boldsymbol{r}) = \frac{1}{4\pi} \int_{\mathbb{R}} \left[ \lim_{y' \to \infty} \frac{\Phi(x',y',0)}{|\boldsymbol{r}-\boldsymbol{r}'|} - \lim_{y' \to -\infty} \frac{\Phi(x',y',0)}{|\boldsymbol{r}-\boldsymbol{r}'|} - \int_{\mathbb{R}}  \Phi(x',y',0)   \frac{y-y'}{|\boldsymbol{r}-\boldsymbol{r}'|^3} dy' \right]dx' .
\]
Now, the Dirichlet boundary condition states that
\[
\Phi(x',y',0) = V_o \hspace{0.5cm}\textbf{if}\hspace{0.5cm}(x',y',0) \in \mathcal{A}_{-} \hspace{0.5cm}\mbox{and}\hspace{0.5cm}\Phi(x',y',0) = 0 \hspace{0.5cm}\textbf{if}\hspace{0.5cm}(x',y',0) \in \mathcal{A}_{+}.
\]
In the gap the scalar electric potential is unknown. Therefore
\[
\Omega_x(\boldsymbol{r}) = - \frac{V_o}{4\pi} \int_{\mathcal{A}_{-}} \frac{y-y'}{|\boldsymbol{r}-\boldsymbol{r}'|^3} dx'dy' - \frac{1}{4\pi} \int_{\mathcal{G}} \Phi(x',y',0) \frac{y-y'}{|\boldsymbol{r}-\boldsymbol{r}'|^3} dx'dy' 
\]
this is
\[
\Omega_x(\boldsymbol{r}) = - \frac{V_o}{4\pi} \int_{\mathcal{A}_{-}}\frac{\partial}{\partial y'}  \left(\frac{1}{|\boldsymbol{r}-\boldsymbol{r}'|}\right) dx'dy' - \frac{1}{4\pi} \int_{\mathcal{G}} \Phi_{\mathcal{G}}(x',y') \frac{y-y'}{|\boldsymbol{r}-\boldsymbol{r}'|^3} dx'dy' \hspace{0.5cm}\mbox{since}\hspace{0.5cm}\frac{\partial}{\partial y'}  \left(\frac{1}{|\boldsymbol{r}-\boldsymbol{r}'|}\right)  =   \frac{y-y'}{|\boldsymbol{r}-\boldsymbol{r}'|^3}.
\]
With $\Phi_{\mathcal{G}}(x',y')$ the scalar electric potential in $\mathcal{G}$. A similar procedure can be used to simplify $\Omega_x(\boldsymbol{r})$, the result is 
\[
\Omega_y(\boldsymbol{r}) =  \frac{V_o}{4\pi} \int_{\mathcal{A}_{-}} \frac{\partial}{\partial x'}  \left(\frac{1}{|\boldsymbol{r}-\boldsymbol{r}'|}\right) dx'dy' + \frac{1}{4\pi} \int_{\mathcal{G}} \Phi_{\mathcal{G}}(x',y') \frac{x-x'}{|\boldsymbol{r}-\boldsymbol{r}'|^3} dx'dy'
\]
The vector field $\boldsymbol{\Omega}(\boldsymbol{r})$ takes the form
\[
\boldsymbol{\Omega}(\boldsymbol{r}) = \Omega_x(\boldsymbol{r})\hat{x} + \Omega_y(\boldsymbol{r})\hat{y} =  \frac{V_o}{4\pi} \int_{\mathcal{A}_{-}} \boldsymbol{\xi}(\boldsymbol{r},\boldsymbol{r}')   d^2\boldsymbol{r}' + \frac{1}{4\pi} \int_{\mathcal{G}} \Phi_{\mathcal{G}}(\boldsymbol{r}') \boldsymbol{\xi}(\boldsymbol{r},\boldsymbol{r}') d^2\boldsymbol{r}'
\]
with
\[
\boldsymbol{\xi}(\boldsymbol{r},\boldsymbol{r}') := \left\{ - \hat{x} \frac{\partial}{\partial y'}  \left(\frac{1}{|\boldsymbol{r}-\boldsymbol{r}'|}\right) + \hat{y} \frac{\partial}{\partial x'}  \left(\frac{1}{|\boldsymbol{r}-\boldsymbol{r}'|}\right) \right\}
\]
which can be written as follows
\[
\boldsymbol{\Omega}(\boldsymbol{r}) =  \frac{V_o}{4\pi} \int_{\mathcal{A}_{-}} \left\{ \frac{\partial}{\partial x'}  \left(\frac{\hat{y}}{|\boldsymbol{r}-\boldsymbol{r}'|}\right) -  \frac{\partial}{\partial y'}  \left(\frac{\hat{x}}{|\boldsymbol{r}-\boldsymbol{r}'|}\right)  \right\}  dx'dy' + \frac{1}{4\pi} \int_{\mathcal{G}} \Phi_{\mathcal{G}}(\boldsymbol{r}') \boldsymbol{\xi}(\boldsymbol{r},\boldsymbol{r}') d^2\boldsymbol{r}'
\]
The first term in the right of the previous equation can be simplified by using the the Green's theorem which states that 
\begin{equation}
\int_{\mathcal{A}} \left[\frac{\partial}{\partial x} M(x,y) - \frac{\partial}{\partial y} L(x,y) \right]dxdy = \rcirclerightint_{\partial \mathcal{A}} L(x,y) dx + M(x,y) dy    
\label{greensTheoremEq}
\end{equation}
for a flat region $\mathcal{A}$ bounded by $\partial \mathcal{A}$ oriented positively. Identifying $M(x,y)$ and $L(x,y)$ as follows 
\[
M \longrightarrow \frac{\hat{y}}{|\boldsymbol{r}-\boldsymbol{r}'|} \hspace{0.5cm} \mbox{and} \hspace{0.5cm} L \longrightarrow \frac{\hat{x}}{|\boldsymbol{r}-\boldsymbol{r}'|}
\]
then
\[
\boldsymbol{\Omega}(\boldsymbol{r}) = \frac{V_o}{4\pi} \oint_{\partial \mathcal{A}_{-}} \frac{\hat{x}}{|\boldsymbol{r}-\boldsymbol{r}'|} dx' + \frac{\hat{y}}{|\boldsymbol{r}-\boldsymbol{r}'|} dy' + \frac{1}{4\pi} \int_{\mathcal{G}} \Phi_{\mathcal{G}}(\boldsymbol{r}') \boldsymbol{\xi}(\boldsymbol{r},\boldsymbol{r}') d^2\boldsymbol{r}'
\]
or
\[
\boldsymbol{\Omega}(\boldsymbol{r}) = \frac{1}{4\pi} \left[V_o \oint_{\partial \mathcal{A}_{-}} \frac{d\boldsymbol{r}'}{|\boldsymbol{r}-\boldsymbol{r}'|} + \int_{\mathcal{G}} \Phi_{\mathcal{G}}(\boldsymbol{r}') \boldsymbol{\xi}(\boldsymbol{r},\boldsymbol{r}') d^2\boldsymbol{r}'\right].
\]
Hence, the field $\boldsymbol{\Omega}(\boldsymbol{r})$ depends exclusively on the shape of the contour $\partial \mathcal{A}$. It is expected that the scalar field $\zeta(\boldsymbol{r})$ and the vector field $\boldsymbol{\Omega}(\boldsymbol{r})$ should be related to the scalar electric potential $\Phi(\boldsymbol{r})$ and the vector electric potential $\boldsymbol{\Theta}(\boldsymbol{r})$ respectively. We know that both fields are valid to represent the electric field in the free charge region
\[
\boldsymbol{E}(\boldsymbol{r}) = -\nabla \Phi(\boldsymbol{r}) \hspace{1.0cm}\mbox{and}\hspace{1.0cm} \boldsymbol{E}(\boldsymbol{r}) =\nabla \times \boldsymbol{\Theta}(\boldsymbol{r})
\]
adding both equations and dividing by 2, then
\begin{equation}
\boldsymbol{E}(\boldsymbol{r}) = -\nabla \left(\frac{\Phi(\boldsymbol{r})}{2}\right)  + \nabla \times \left(\frac{\boldsymbol{\Theta}(\boldsymbol{r})}{2}\right)  
\label{sumTwoExpressionsEq}    
\end{equation}
Comparing Eqs.~(\ref{sumTwoExpressionsEq}) and (\ref{hemholtzDescompositionEFieldEq})  it is obtained $\zeta(\boldsymbol{r})=\Phi(\boldsymbol{r})/2$ and $\boldsymbol{\Omega}(\boldsymbol{r})=\boldsymbol{\Theta}(\boldsymbol{r})/2$\footnote{In principle, this type of identification among potentials is not unique since we can define them as follows 
\[
\zeta(\boldsymbol{r})=\frac{\gamma_1}{\gamma_1+\gamma_2}\Phi(\boldsymbol{r}) \hspace{0.5cm} \mbox{and} \hspace{0.5cm} \boldsymbol{\Omega}(\boldsymbol{r})=\frac{\gamma_2}{\gamma_1+\gamma_2}\boldsymbol{\Theta}(\boldsymbol{r})  
\]
with $\gamma_1,\gamma_2 \in \mathbb{R}$ such that $\gamma_1+\gamma_2 \neq 0$. In this document, we choose $\gamma_1=\gamma_2=1$ without loss of generality. 

}, then the electric scalar and vector potential are
\begin{equation}
    \Phi( \boldsymbol{r})  =
 2 \zeta( \boldsymbol{r})  = \frac{1}{4\pi\epsilon_o} \int_{\mathbb{R}^2} \frac{\sigma(x',y')}{|\boldsymbol{r}-\boldsymbol{r}'|} dx'dy'
 \label{scalarPotentiaEq}
\end{equation}
and
\begin{equation}
\boxed{
\boldsymbol{\Theta}(\boldsymbol{r}) = \frac{1}{2\pi} \left[V_o\rcirclerightint_{\partial \mathcal{A}_{-}} \frac{d\boldsymbol{r}'}{|\boldsymbol{r}-\boldsymbol{r}'|} + \int_{\mathcal{G}} \Phi_{\mathcal{G}}(\boldsymbol{r}') \boldsymbol{\xi}(\boldsymbol{r},\boldsymbol{r}') d^2\boldsymbol{r}'\right]     
} \hspace{0.5cm} \mbox{Electric vector potential.}
\label{electricVectorPotentiaEq}
\end{equation}
where the path of integration along $\partial \mathcal{A}$ was taken counter-clockwise. Note that Eq.~(\ref{scalarPotentiaEq}) is the usual integral expression of the potential scalar potential obtained from the Coulomb's law
\[
\Phi( \boldsymbol{r})  = \frac{1}{4\pi\epsilon_o} \int_{\mathbb{R}^3} \frac{\rho(\boldsymbol{r}')}{|\boldsymbol{r}-\boldsymbol{r}'|} d^3 \boldsymbol{r}'
\]
where volume charge density is $\rho(\boldsymbol{r}') = \sigma(x',y')\delta(z')$. Unfortunately, Eq.~(\ref{scalarPotentiaEq}) is not very useful in the current problem since the surface charge density $\sigma(x',y')$ is unknown. On the other hand, the electric vector potential results advantageous since it can be evaluated without having the surface charge density in advance.   

\section{The Gapless Surface Electrode GSE}
In this particular case the thickness of the gap is zero $\nu=0$, and the electric vector potential in Eq.~(\ref{electricVectorPotentiaEq}) takes the form
\begin{equation}
\boldsymbol{\Theta}^{\mbox{\tiny \textbf{GSE}}}(\boldsymbol{r}) = \frac{V_o}{2\pi} \rcirclerightint_{\partial \mathcal{A}_{-}} \frac{d\boldsymbol{r}'}{|\boldsymbol{r}-\boldsymbol{r}'|}      
\label{electricVectorPotentiaGaplessEq}
\end{equation}
we may use Eq.~(\ref{electricVectorPotentiaGaplessEq}) to obtain expressions for the electric field an the surface charge density of the GSE.
\subsection{Electric field}
\label{ElectricFieldSection} 
In this subsection we shall find  the electric field from the electric vector potential. By definition Eq.~(\ref{electricVectorPotentiaEq})  the electric field is 
\[
\boldsymbol{E}^{\mbox{\tiny \textbf{GSE}}}(\boldsymbol{r}) = \nabla \times \boldsymbol{\Theta}^{\mbox{\tiny \textbf{GSE}}}(\boldsymbol{r}) = \frac{V_o}{2\pi} \rcirclerightint_{\partial \mathcal{A}} \nabla \times \frac{d\boldsymbol{r}'}{|\boldsymbol{r}-\boldsymbol{r}'|}
\]
we may use the following vector calculus identity
\begin{equation}
\nabla \times [\psi(\boldsymbol{r}) \times \boldsymbol{F}(\boldsymbol{r})] = \psi(\boldsymbol{r}) \nabla \times \boldsymbol{F}(\boldsymbol{r}) + \nabla \psi(\boldsymbol{r}) \times \boldsymbol{F}(\boldsymbol{r})       
\label{vectorCalculusIdentityCurlEq}
\end{equation}
choosing $\psi$ and $\boldsymbol{F}$ as $|\boldsymbol{r}-\boldsymbol{r}'|^{-1}$ and $d\boldsymbol{r}'$ respectively, we may write
\[
\nabla \times \frac{d\boldsymbol{r}'}{|\boldsymbol{r}-\boldsymbol{r}'|} = \nabla \left( \frac{1}{|\boldsymbol{r}-\boldsymbol{r}'|} \right) \times d\boldsymbol{r}' = d\boldsymbol{r}' \times \frac{\boldsymbol{r}-\boldsymbol{r}'}{|\boldsymbol{r}-\boldsymbol{r}'|^3}. 
\]
Hence the electric field takes the form
\begin{equation}
\boxed{
\boldsymbol{E}^{\mbox{\tiny \textbf{GSE}}}(\boldsymbol{r}) = \frac{V_o}{2\pi} \textbf{\mbox{sgn}}(z)  \rcirclerightint_{\partial \mathcal{A}}   \frac{d\boldsymbol{r}' \times (\boldsymbol{r}-\boldsymbol{r}')}{|\boldsymbol{r}-\boldsymbol{r}'|^3} 
}
\label{BSLElectricPontentialFromThetaEq}
\end{equation}
where the path of integration is counter-clockwise. It is also possible to find this formula from the scalar electric potential $\Phi(\boldsymbol{r})$ or employing analogies between the GSE and magnetostatics (see the Appendix Section \ref{ElectricFieldFromTheScalarPotentialSection}).

\subsection{Charge density}
\subsubsection{Circular GSE}
\begin{figure}[H]
\centering %system.pdf
\includegraphics[width=0.45\textwidth]{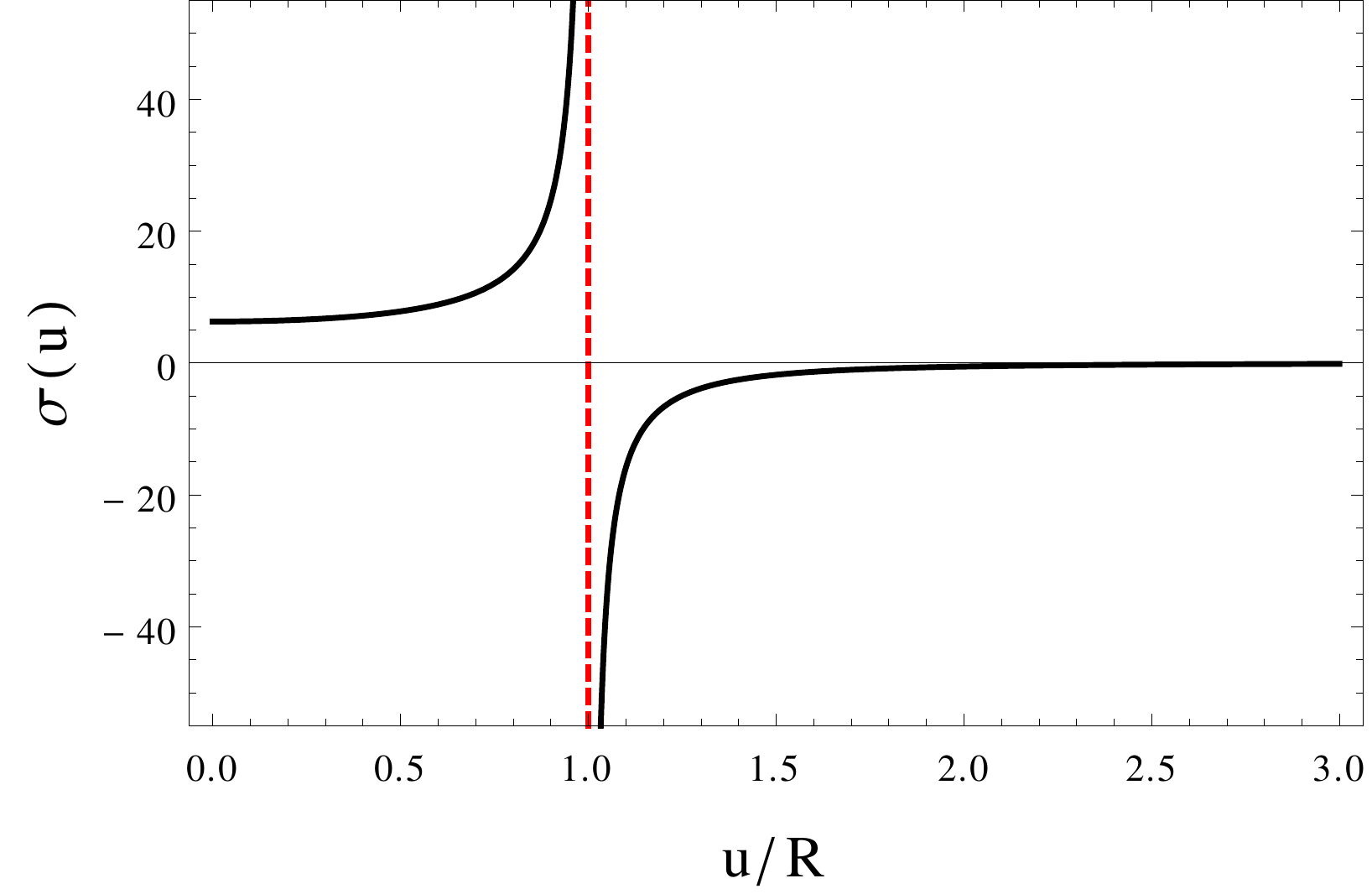}
\includegraphics[width=0.5\textwidth]{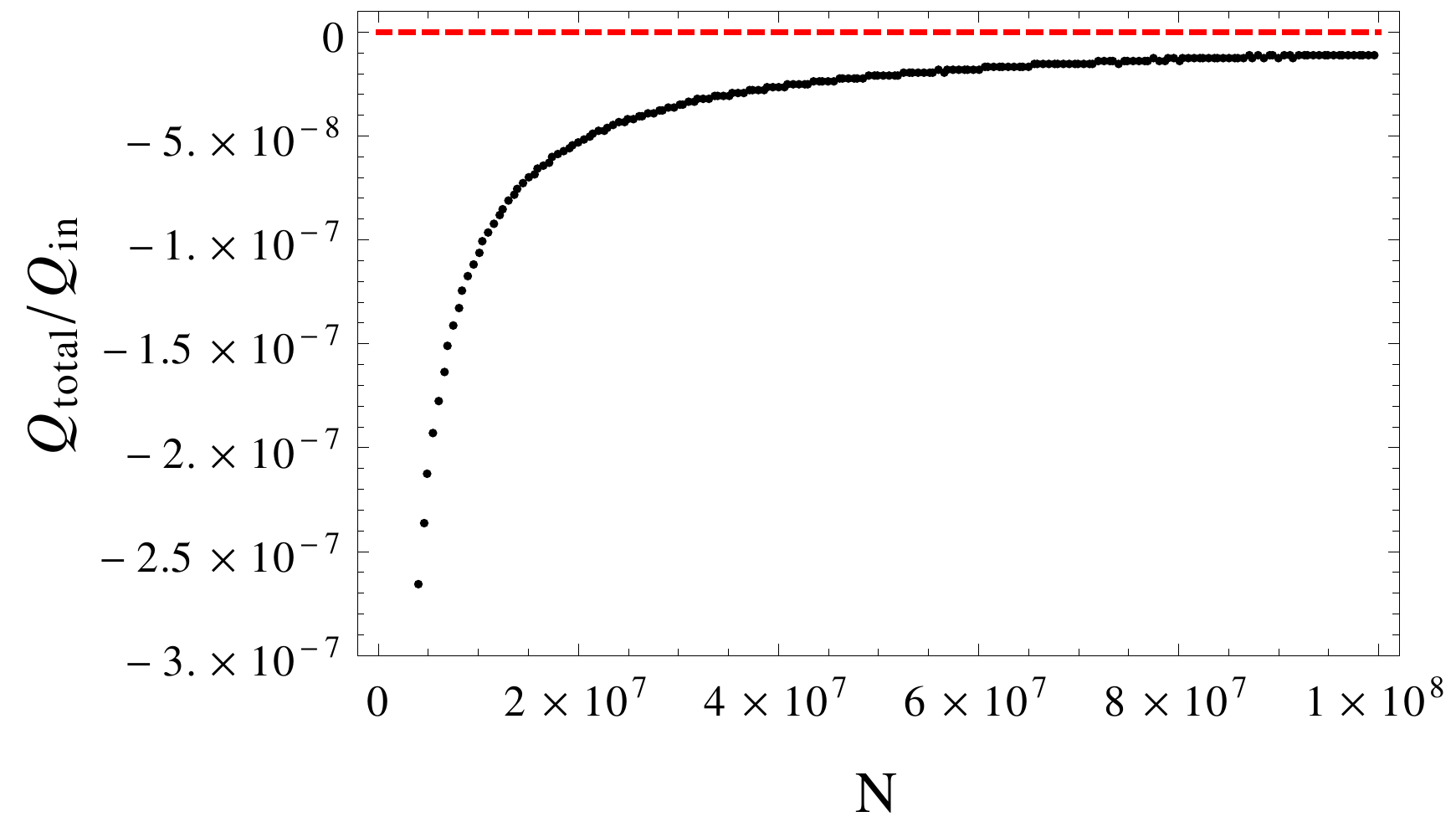}
    \caption[Circular SE.]{Circular SE. (left) Surface charge density accordig to Eq.~(\ref{surfaceChargeDensityCircularSEEq}). (right) Charge ratio.}
\label{chargeDensityFig}
\end{figure}

In this section, we shall use the vector electric potential to calculate the surface density charge of the circular SE. If $\partial\mathcal{A}$ is a circular path of radius $R$ on the $xy$-plane and centered at the origin, then the electric vector potential in Eq.~(\ref{electricVectorPotentiaEq}) takes the form
\[
\boldsymbol{\Theta}^{\mbox{\tiny \textbf{GSE}}}(\boldsymbol{r}) = \frac{V_o R}{2\pi} \int_{0}^{2\pi} \frac{\cos(\phi-\phi')d\phi'}{|\boldsymbol{r}-\boldsymbol{r}'|} \hat{\phi}(\boldsymbol{r})
\]
where $|\boldsymbol{r}-\boldsymbol{r}'|=\sqrt{r^2+R^2-2rR\cos(\phi-\phi')\sin\theta}$. The previous integral is independent of $\phi$ since SE is an axially symmetric system, and it can be written as follows
\begin{equation}
\Theta_{\phi}^{\mbox{\tiny \textbf{GSE}}}(r,\theta) = \frac{V_o}{2\pi}\frac{4R}{\sqrt{R^2+r^2+2Rr\sin\theta}}\left[\frac{(2-\gamma^2)K(\gamma^2)-2\bar{E}(\gamma^2)}{\gamma^2}\right]
\label{ThetaPhiGaplessEq}
\end{equation}
where $\gamma^2(r,\theta):=4Rr\sin\theta / (R^2+r^2+2Rr\sin\theta)$, $K$ and $\bar{E}$ are the complete elliptic integrals of first and second kind respectively. According to Eq.~(\ref{surfaceChargeDensityLimitEq}) the charge density can be written as follows
\begin{equation}
\sigma^{\mbox{\tiny \textbf{GSE}}}(u) = \frac{2 \epsilon_o}{u} \lim_{\theta \to \frac{\pi}{2}} \frac{\partial}{\partial r} \left[r \Theta_{\phi}(r,\theta)\right]
\end{equation}
where $u=\sqrt{x^2+y^2}$, thus evaluating the limit it is found
\begin{equation}
\sigma^{\mbox{\tiny \textbf{GSE}}}(u) = \frac{2\epsilon_o V_o}{\pi}\left[ \frac{1}{R-u}\underbar{E}\left(\frac{4Ru}{(R+u)^2}\right) + \frac{1}{R+u} K\left(\frac{4Ru}{(R+u)^2}\right)\right].
\label{surfaceChargeDensityCircularSEEq}
\end{equation}
The surface charge density has a discontinuity at $u=R$ since
\[
\lim_{u \to R^{-}} \sigma(u) \rightarrow \infty  \hspace{0.5cm}\mbox{and}\hspace{0.5cm} \lim_{u \to R^{+}} \sigma(u) \rightarrow - \infty
\]
as it is shown in Fig.~\ref{chargeDensityFig}-left. This implies that charge into region $\mathcal{A}$ is positive, and negative outside it.

The total charge of the circular GSE is given by
\[
Q_{total} = 2\pi\int_{\mathbb{R}^+}\sigma(u)udu = \lim_{N\to\infty}\left[Q_{in}(N) + Q_{out}(N)\right]
\]
where
\[
Q_{in}(N) = 2\pi\int_{0}^{(1-1/N)R}\sigma(u) u du  \hspace{0.5cm}\mbox{and}\hspace{0.5cm}Q_{out}(N) = 2\pi\int_{(1+1/N)R}^{\infty}\sigma(u) u du.  
\]
These integrals are difficult to evaluate as $N\rightarrow\infty$, however it is possible to explore numerically the quantity $Q_{in}(N)+Q_{out}(N)$ which approaches to the total charge $Q_{total}$ as $N$ is increased. A plot of $1+Q_{in}(N)/Q_{out}(N)$ as a function of $N$ is shown in Fig.~\ref{chargeDensityFig}-right. Since $1+Q_{in}(N)/Q_{out}(N)$ tends to zero as $N$ is increased, then it tell us that system should be globally neutral $Q_{total}=0$. The electric field is obtained from  the curl of the electric vector potential, the result for $z>0$ is  
\begin{equation}
 \boldsymbol{E}^{\mbox{\tiny \textbf{GSE}}}_{r}(\boldsymbol{r}) = (\nabla\times\boldsymbol{\Theta}^{\mbox{\tiny \textbf{GSE}}})_r = \frac{R^2 V_1}{2\pi} \frac{4\cos\theta}{\mathscr{r}_{-}(r,\theta)^2 \mathscr{r}_{+}(r,\theta)} \underbar{E}\left[ \frac{4rR\sin\theta}{ \mathscr{r}_{+}(r,\theta)^2 }\right],
    \label{ErGaplessEq}
\end{equation}

\begin{equation}
\boldsymbol{E}^{\mbox{\tiny \textbf{GSE}}}_{\theta}(\boldsymbol{r}) = (\nabla\times\boldsymbol{\Theta}^{\mbox{\tiny \textbf{GSE}}})_\theta  = \frac{V_1}{\pi} \frac{\csc\theta}{\mathscr{r}_{-}^2 \mathscr{r}_{+}} \left[  (r^2+R^2\cos(2\theta))\underbar{E}\left( \frac{4rR\sin\theta}{ \mathscr{r}_{+}^2 }\right) - \mathscr{r}_{-}^2 K\left( \frac{4rR\sin\theta}{ \mathscr{r}_{+}^2 }\right)\right]
    \label{EThetaGaplessEq}
\end{equation}
and $\boldsymbol{E}^{\mbox{\tiny \textbf{GSE}}}_{\phi}(\boldsymbol{r})=0$ where it is defined $\mathscr{r}_{\pm}(r,\theta) = \sqrt{r^2 + R^2 \pm 2rR\sin\theta}$. 

\subsection{Polygonal interconnected GSE}
Let us define a contour $\partial \mathcal{A}=\bigcup_{n=1}^N\overline{P_{n}P_{n+1}}$ composed by rectilinear segments connecting a set of points $P_1,\ldots,P_{N_S}$ on the plane $z=0$, with $P_{N_S+1}=P_{1}$ and $N_S$ total number of segments. We assume that polygon vertices are numbered in a counter clockwise sense. Let us suppose that set $\{P_n\}_{1 \leq n \leq N_S}$ lie on a curve whose polar equation is known $\mathscr{R}=\mathscr{R}(\phi)$. The length of arc of the corresponding polygon can be written as folows 
\[
s(\phi) = s^{(n+1)}(\phi) + \mathcal{L}_n ,
\]
with $\mathcal{L}_n = \sum_{j=1}^{n} l_j$, $l_n$ the length of $\overline{P_{n}P_{n+1}}$ and $s^{(n)}(\phi) \in [0, l_{n}]$. The electric vector potential of the $n$th-segment is 
\[
\Theta_n^{\mbox{\tiny \textbf{GSE}}}(\boldsymbol{r}) = \frac{V_o}{2\pi} \int_{ \partial \mathcal{A}_{n,n+1} }  \frac{d\boldsymbol{s}^{(n+1)}}{|\boldsymbol{r}-s^{(n+1)}\hat{t}_{n}|},
\]
with $\hat{t}_{n}:=P_{n+1}-P_{n}$ the tangent vector. The total electric vector potential field is obtained by superposing the electric vector potentials of each segment. If $\Theta^{\mbox{\tiny \textbf{GSE}}}_n(\boldsymbol{r})$ is solved in the reference frame centered at $P_n$ defined by the unit vectors $\hat{t}_n \times \hat{e}_3$ and $\hat{t}_n$ then
\[
\boldsymbol{\Theta}^{\mbox{\tiny \textbf{GSE}}}(\boldsymbol{r}) = \sum_{n=1}^{N_S} R_z(\gamma_n)\boldsymbol{\Theta}^{\mbox{\tiny \textbf{GSE}}}_{n}(\hat{e}_1 \cdot R_z(\gamma_n)^{T}(\boldsymbol{r}-P_n), \hat{e}_2 \cdot R_z(\gamma_n)^{T}(\boldsymbol{r}-P_n), \hat{e}_3 \cdot R_z(\gamma_n)^{T}(\boldsymbol{r}-P_n)) 
\]
where $R_z(\gamma)$ is a counter-clock wise rotation matrix around the z-axis, $\hat{e}_i$ the Cartesian unit vectors, and $\gamma_n$ the angle between the $x$-axis and the $n$-th tangent vector $\hat{t}_n$. The surface charge density is 

\begin{equation}
\sigma^{\mbox{\tiny \textbf{GSE}}}(u,\phi) = \frac{2 \epsilon_o}{u} \lim_{\theta \to \frac{\pi}{2}^{+}} \left\{\frac{\partial}{\partial r} \left[r \Theta^{\mbox{\tiny \textbf{GSE}}}_{\phi}(r,\theta,\phi)\right] - \frac{\partial}{\partial \phi} \left[\Theta^{\mbox{\tiny \textbf{GSE}}}_{r}(r,\theta,\phi)\right]\right\}
\label{surfaceChargeDensityLimitGenericSEEq}
\end{equation}
this is reduced to
\begin{equation}
\sigma^{\mbox{\tiny \textbf{GSE}}}(x,y) =\frac{\epsilon_o V_o}{\pi}\sum_{n=1}^{N_S} \frac{(D_z)_n(\boldsymbol{r})}{\tilde{\lambda}_n(\boldsymbol{r})^2-\Lambda_n(\boldsymbol{r})^2} \left[\frac{l_n-\Lambda_n(\boldsymbol{r})}{\sqrt{\tilde{\lambda}_n(\boldsymbol{r})^2+l_n^2 - 2 l_n \Lambda_n(\boldsymbol{r})}}+ \frac{\Lambda_n(\boldsymbol{r})}{\tilde{\lambda}_n(\boldsymbol{r})}\right],
\label{surfaceChargeDensityAnalyticPolyEq}
\end{equation}
when a polygonal boundary is considered, where 
\[
\tilde{\lambda}_n(\boldsymbol{r}) = \sqrt{x^2+y^2 + \mathscr{R}_n[\mathscr{R}_n - 2(x\cos\gamma_n+y\sin\gamma_n)] },\hspace{0.5cm}\Lambda_n(\boldsymbol{r}) = y\cos\gamma_n - x\sin\gamma_n + \mathscr{R}_n \sin(\gamma_n-\beta_n),
\]
$(D_z)_n = -(x-x_n)\cos\gamma_n+(y-y_n)\sin\gamma_n$, $\beta_n=2n\pi/N$ and $\mathscr{R}_n=\mathscr{R}(\gamma_n)$. The total charge of the SE can be obtained from
\[
Q_{total} = \int_{\mathbb{R}^2} \sigma(\boldsymbol{r})d^2\boldsymbol{r}.
\]
As occurs with the circular GSE, the previous integral is difficult to solve since the surface charge density diverges at $\partial \mathcal{A}$. However, we may expect that $Q_{total}$ should be zero since the electric field is a free-divergence field for $z \neq 0$, then it behaves as it would be solenoidal field in $\mathcal{D}$\footnote{Take into account that the electric field of the planar SE is not a fully solenoidal field in the $\mathbb{R}^3$ space due to its reflection symmetry with respect the plane $z=0$.}. Therefore, each flow line of $\boldsymbol{E}(\boldsymbol{r})$ that leaves region $\mathcal{A}$ must enter to the region region $\mathbb{R}^2 \mathbin{\backslash} \mathcal{A}$ if $V_o>0$. This implies that the total electric field flow through a surface $S$ containing the $xy$-plane would be zero as well as the total charge inside such surface.

\section{The Gapped SE}

\subsection{Gapped Circular SE}
In this subsection we shall consider a gapped circular SE. The gap is the annular region $\mathscr{G}=\textbf{ann}(R_{-}, R_{+})$ between the concentric circles of radius
$R_{-}=R-\nu/2$ and $R_{+}=R+\nu/2$. The electric vector potential is
\[
\boldsymbol{\Theta}(\boldsymbol{r}) = \frac{1}{2\pi} \left[V_o\rcirclerightint_{\partial \mathcal{A}_{-}} \frac{d\boldsymbol{r}'}{|\boldsymbol{r}-\boldsymbol{r}'|} + \frac{1}{2\pi}\int_{\mathbf{ann}(R_{-}, R_{+})} \Phi_{\mathcal{G}}(\boldsymbol{r}') \boldsymbol{\xi}(\boldsymbol{r},\boldsymbol{r}') d^2\boldsymbol{r}'\right].     
\]
Since this problem is axially symmetric, the scalar electric potential in the gap $\Phi_{\mathcal{G}}=\Phi_{\mathcal{G}}(u)$ depends only on the radial coordinate on the plane. The gap can be written as follows
\[
\mathscr{G} = \mathscr{G}_1 \cup \mathscr{G}_2 \cup \ldots \cup \mathscr{G}_N = \bigcup_{m=1}^N \mathscr{G}_m
\]
with $\mathscr{G}_m=\textbf{ann}(u_{m-1}, u_m)$, $u_o=R_{-}$, $u_N=R_{+}$ and $\left\{u_m\right\}_{m=0,\ldots,N}$ a set under the condition $u_m<u_{m+1} \forall m \in [0,N]$. Therefore
\[
\boldsymbol{\Theta}(\boldsymbol{r}) = \frac{V_o}{2\pi} \rcirclerightint_{\partial \mathcal{A}_{-}} \frac{d\boldsymbol{r}'}{|\boldsymbol{r}-\boldsymbol{r}'|} + \lim_{N\rightarrow\infty} \sum_{n=0}^N \Phi_{\mathcal{G}m} \int_{\mathbf{ann}(u_{m-1}, u_m)}  \boldsymbol{\xi}(\boldsymbol{r},\boldsymbol{r}') d^2\boldsymbol{r}'    
\]
with $\Phi_{\mathcal{G}}m$ the average value of the scalar electric potential in $\mathscr{G}_m$.

\begin{figure}[H]
\centering %system.pdf
\includegraphics[width=0.6\textwidth]{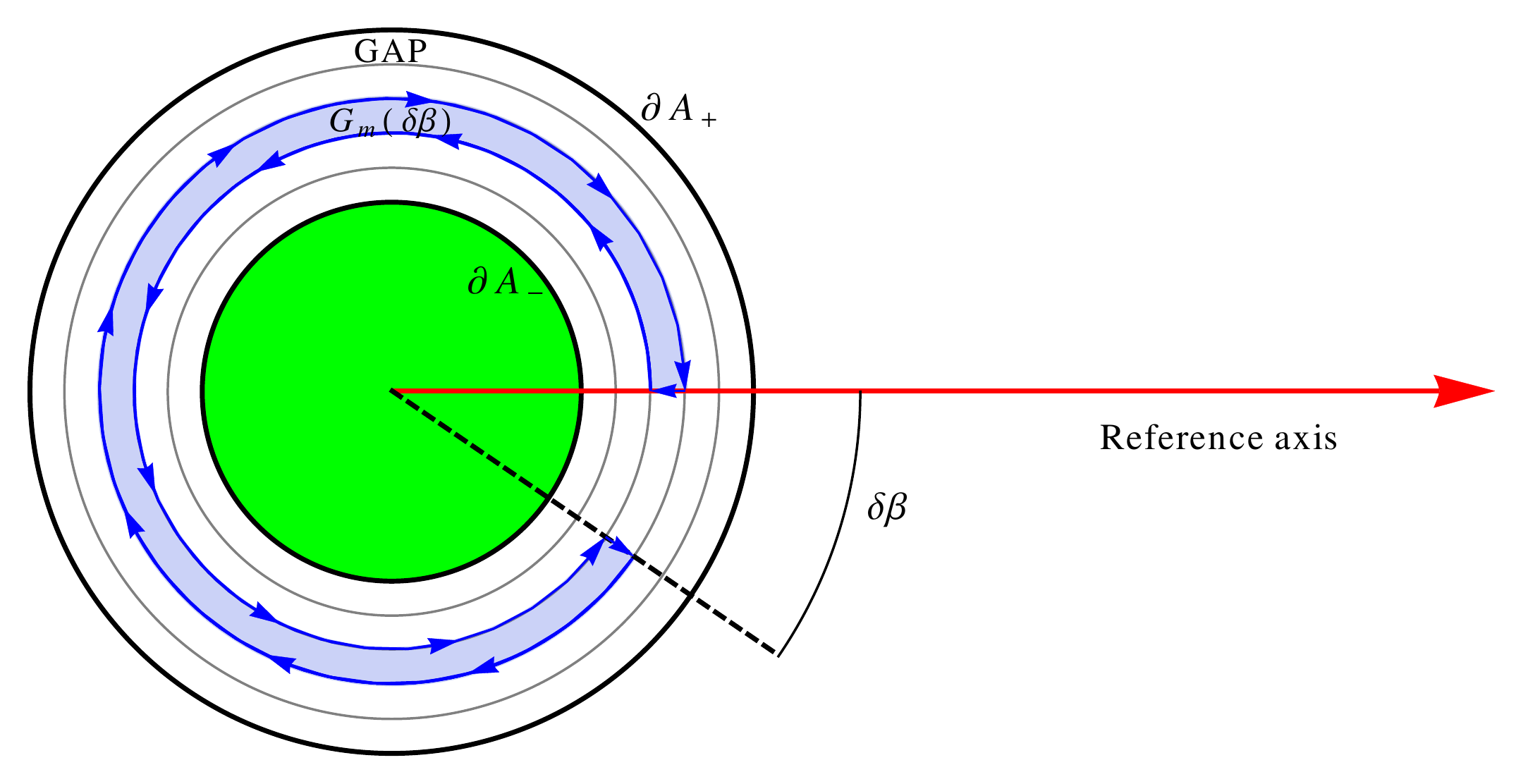}
    \caption[The region $G_m(\delta\beta)$.]{The region $G_m(\delta\beta)$.}
\label{theRegionCircularSEFig}
\end{figure}

It is possible to use the Green's theorem as we did in Section \ref{HelmholtzDecompositionTheoremSectionLabel} to write
\[
\int_{\mathbf{ann}(u_{m-1}, u_m)}  \boldsymbol{\xi}(\boldsymbol{r},\boldsymbol{r}') d^2\boldsymbol{r}' =  \oint_{\partial \mathcal{G}_{m,m+1}} \frac{d\boldsymbol{r}'}{|\boldsymbol{r}-\boldsymbol{r}'|}
\]
where 
\[
\partial\mathcal{G}_{m,m+1} = \lim_{\delta \beta \rightarrow 0} \partial G_m(\delta\beta) = \lim_{\delta \beta \rightarrow 0} \partial\mathcal{G}_{m}^+(\delta\beta) \cup l_m^{-}(\delta\beta) \cup \partial\mathcal{G}_{m}^-(\delta\beta) \cup l_m^{+}(\delta\beta).   
\]
The region $G_{m}(\beta)$ is represented in Fig.~\ref{theRegionCircularSEFig} and its contour $\partial G_m(\delta\beta)$ is the union (choosing the y-axis as reference axis) of the following trajectories
\[
\partial\mathcal{G}_{m}^+(\delta\beta) = \{ (u_{m},\phi,0) \hspace{0.25cm} : \hspace{0.25cm} \phi \in [0,2\pi-\delta\beta] \} 
\]
in a counterclockwise sense
\[
\partial\mathcal{G}_{m}^-(\delta\beta) = \{ (u_{m-1},\phi,0) \hspace{0.25cm} : \hspace{0.25cm} \phi \in [\delta\beta,2\pi] \}
\]
in a clockwise sense
\[
l_m^{+} = \{ (u,0,0) \hspace{0.25cm} : \hspace{0.25cm} u \in [u_{m-1},u_{m}] \}
\]
it is a radially drawn path from the inside to outside, and 
\[
l_m^{-}(\delta\beta) = \{ (u,\delta\beta,0) \hspace{0.25cm} : \hspace{0.25cm} u \in [u_{m-1},u_{m}] \}
\]
it is a radially drawn path from outside to inside. The electric vector potential takes the form
\[
\boldsymbol{\Theta}(\boldsymbol{r}) = \frac{V_o}{2\pi} \rcirclerightint_{\partial \mathcal{A}_{-}} \frac{d\boldsymbol{r}'}{|\boldsymbol{r}-\boldsymbol{r}'|} + \lim_{N\rightarrow\infty} \sum_{m=1}^N \frac{\Phi_{\mathcal{G}m}}{2\pi} \left\{  \lcirclerightint_{\partial \mathcal{G}_{m}^{-}(0)} \frac{d\boldsymbol{r}'}{|\boldsymbol{r}-\boldsymbol{r}'|} + \rcirclerightint_{\partial \mathcal{G}_m^{+}(0)} \frac{d\boldsymbol{r}'}{|\boldsymbol{r}-\boldsymbol{r}'|} + \int_{l_m^{+}} \frac{d\boldsymbol{r}'}{|\boldsymbol{r}-\boldsymbol{r}'|} + \lim_{\delta\beta \rightarrow 0}\int_{l_m^{-}(\delta\beta)} \frac{d\boldsymbol{r}'}{|\boldsymbol{r}-\boldsymbol{r}'|} \right\}    
\]
or
\[
\boldsymbol{\Theta}(\boldsymbol{r}) = \frac{V_o}{2\pi} \rcirclerightint_{\partial \mathcal{A}_{-}} \frac{d\boldsymbol{r}'}{|\boldsymbol{r}-\boldsymbol{r}'|} + \frac{1}{2\pi}\lim_{N\rightarrow\infty} \sum_{m=1}^N \Phi_{\mathcal{G}m} \left[  \lcirclerightint_{\partial \mathcal{G}_{m}^{-}(0)} \frac{d\boldsymbol{r}'}{|\boldsymbol{r}-\boldsymbol{r}'|} + \rcirclerightint_{\partial \mathcal{G}_m^{+}(0)} \frac{d\boldsymbol{r}'}{|\boldsymbol{r}-\boldsymbol{r}'|}  \right]    
\]
since $\lim_{\delta\beta \to 0}l_m^{-}(\delta\beta)$ is the reversed trajectory of $l_m^{+}$.
%\[
%\begin{split}
%\boldsymbol{\Theta}(\boldsymbol{r}) = & \frac{V_o}{2\pi} \rcirclerightint_{\partial \mathcal{A}_{-}} \frac{d\boldsymbol{r}'}{|\boldsymbol{r}-\boldsymbol{r}'|} +  \frac{\Phi_{\mathcal{G}1}}{2\pi}  \lcirclerightint_{\partial \mathcal{G}_{1}^{-}(0)} %\frac{d\boldsymbol{r}'}{|\boldsymbol{r}-\boldsymbol{r}'|} + \frac{\Phi_{\mathcal{G}1}}{2\pi} \rcirclerightint_{\partial \mathcal{G}_1^{+}(0)} \frac{d\boldsymbol{r}'}{|\boldsymbol{r}-\boldsymbol{r}'|} + \\ & \left.\frac{\Phi_{\mathcal{G}2}}{2\pi}  %\lcirclerightint_{\partial \mathcal{G}_{2}^{-}(0)} \frac{d\boldsymbol{r}'}{|\boldsymbol{r}-\boldsymbol{r}'|} + \frac{\Phi_{\mathcal{G}2}}{2\pi} \rcirclerightint_{\partial \mathcal{G}_2^{+}(0)} \frac{d\boldsymbol{r}'}{|\boldsymbol{r}-\boldsymbol{r}'|} + \cdots + %\frac{\Phi_{\mathcal{G}N}}{2\pi}  \lcirclerightint_{\partial \mathcal{G}_{N}^{-}(0)} \frac{d\boldsymbol{r}'}{|\boldsymbol{r}-\boldsymbol{r}'|} + \frac{\Phi_{\mathcal{G}N}}{2\pi} \rcirclerightint_{\partial \mathcal{G}_N^{+}(0)} %\frac{d\boldsymbol{r}'}{|\boldsymbol{r}-\boldsymbol{r}'|}\right|_{N\rightarrow\infty}
%\end{split}
%\]
The scalar electric potential on the plane must be continuous, therefore $\Phi_{\mathcal{G}1} = V_o$ and  $\lim_{N \to \infty}\Phi_{\mathcal{G}N} = 0$. Then, first two terms of the right side of the previous equation
\[
\frac{V_o}{2\pi} \rcirclerightint_{\partial \mathcal{A}_{-}} \frac{d\boldsymbol{r}'}{|\boldsymbol{r}-\boldsymbol{r}'|} +  \frac{\Phi_{\mathcal{G}1}}{2\pi}  \lcirclerightint_{\partial \mathcal{G}_{1}^{-}(0)} \frac{d\boldsymbol{r}'}{|\boldsymbol{r}-\boldsymbol{r}'|} = 0
\]
cancel each other since $\mathcal{A}_{-}=\mathcal{G}_{1}^{-}(0)$ and the sense of the line integrals are opposite. On the other hand, $\mathcal{G}_{m+1}^{-}(0) = \mathcal{G}_{m}^{+}(0) \forall m \in [1,N)$ hence

\[
\begin{split}
\boldsymbol{\Theta}(\boldsymbol{r}) = &  \frac{\Phi_{\mathcal{G}1}}{2\pi}  \lcirclerightint_{\partial \mathcal{G}_{1}^{+}(0)} \frac{d\boldsymbol{r}'}{|\boldsymbol{r}-\boldsymbol{r}'|} + \frac{\Phi_{\mathcal{G}2}}{2\pi} \rcirclerightint_{\partial \mathcal{G}_1^{+}(0)} \frac{d\boldsymbol{r}'}{|\boldsymbol{r}-\boldsymbol{r}'|} + \ldots + \\ & \left.\frac{\Phi_{\mathcal{G}N-1}}{2\pi}  \lcirclerightint_{\partial \mathcal{G}_{N-1}^{+}(0)} \frac{d\boldsymbol{r}'}{|\boldsymbol{r}-\boldsymbol{r}'|} + \frac{\Phi_{\mathcal{G}N}}{2\pi} \rcirclerightint_{\partial \mathcal{G}_{N-1}^{+}(0)} \frac{d\boldsymbol{r}'}{|\boldsymbol{r}-\boldsymbol{r}'|} + \frac{\Phi_{\mathcal{G}N}}{2\pi} \rcirclerightint_{\partial \mathcal{G}_N^{+}(0)} \frac{d\boldsymbol{r}'}{|\boldsymbol{r}-\boldsymbol{r}'|}\right|_{N\rightarrow\infty}
\end{split}
\]

\begin{figure}[H]
\centering %system.pdf
\includegraphics[width=0.3\textwidth]{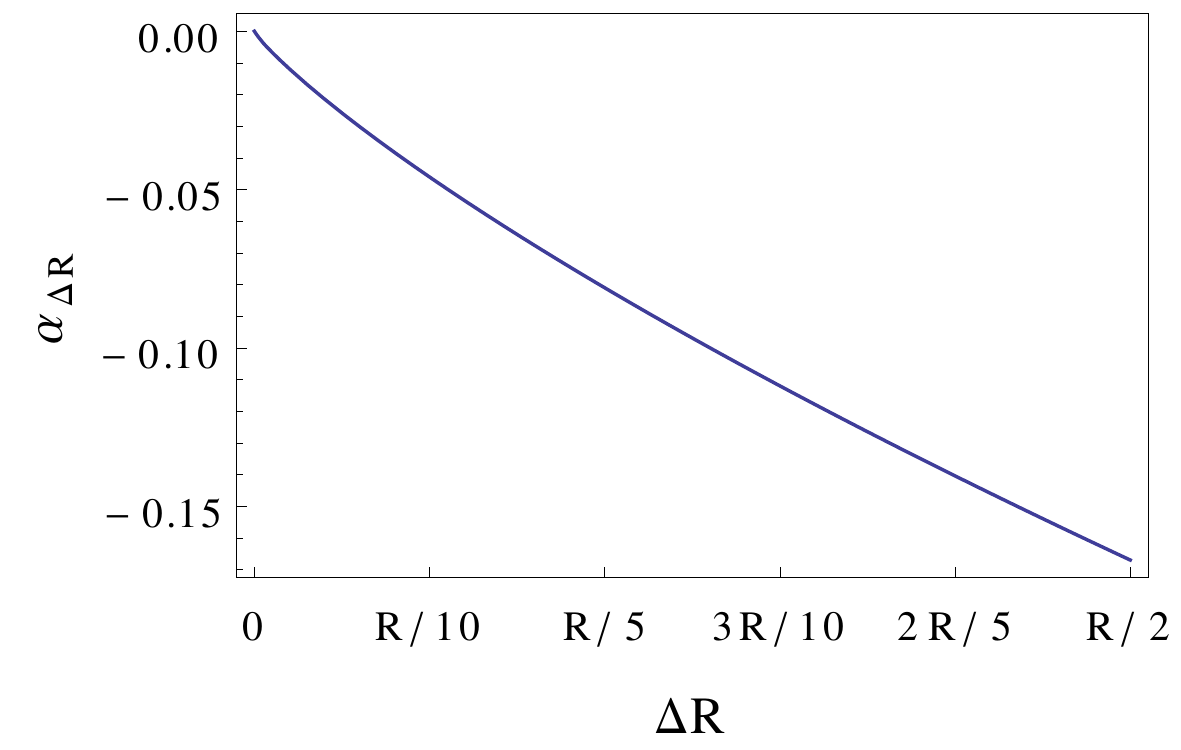}
\includegraphics[width=0.33\textwidth]{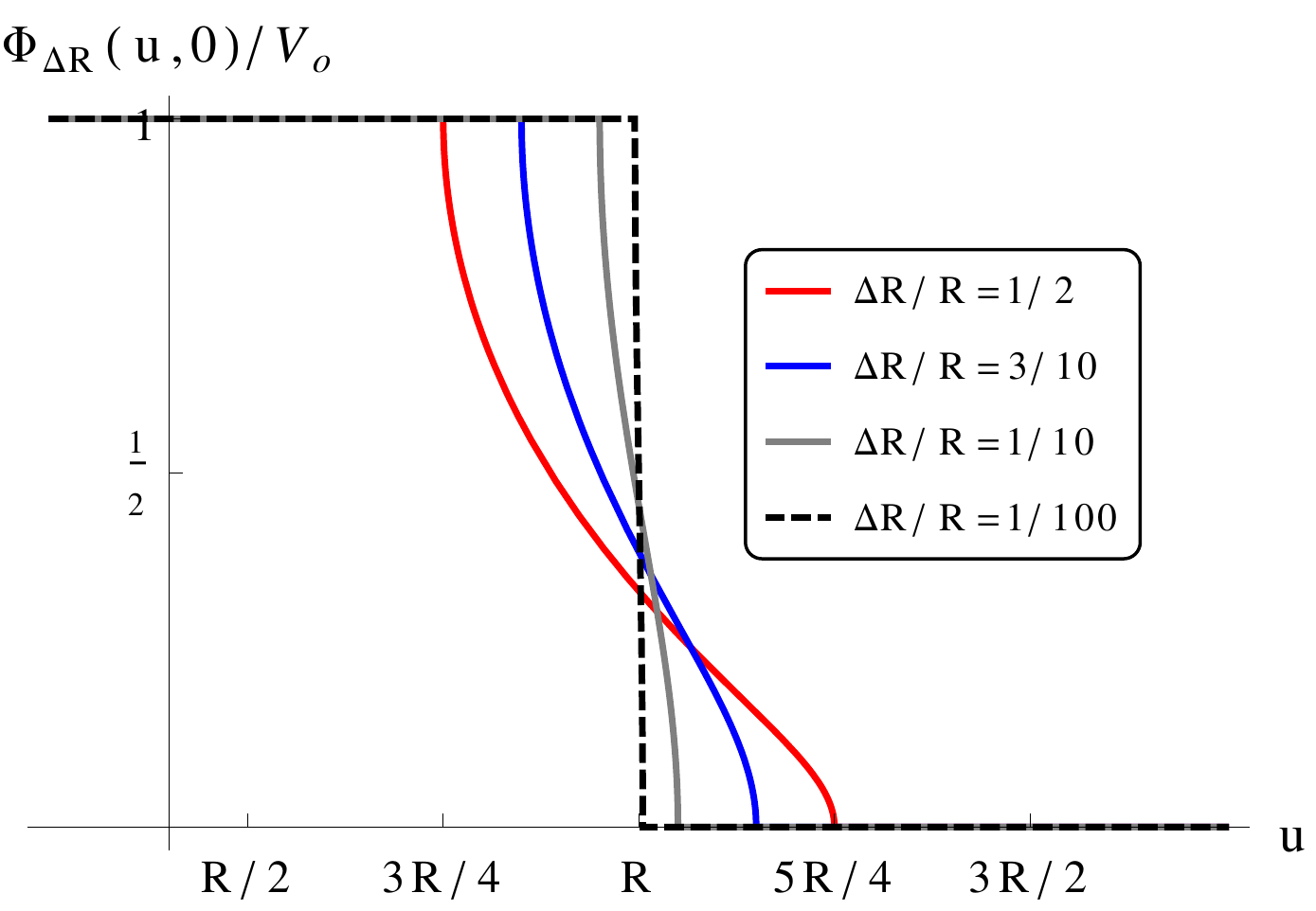}
\includegraphics[width=0.33\textwidth]{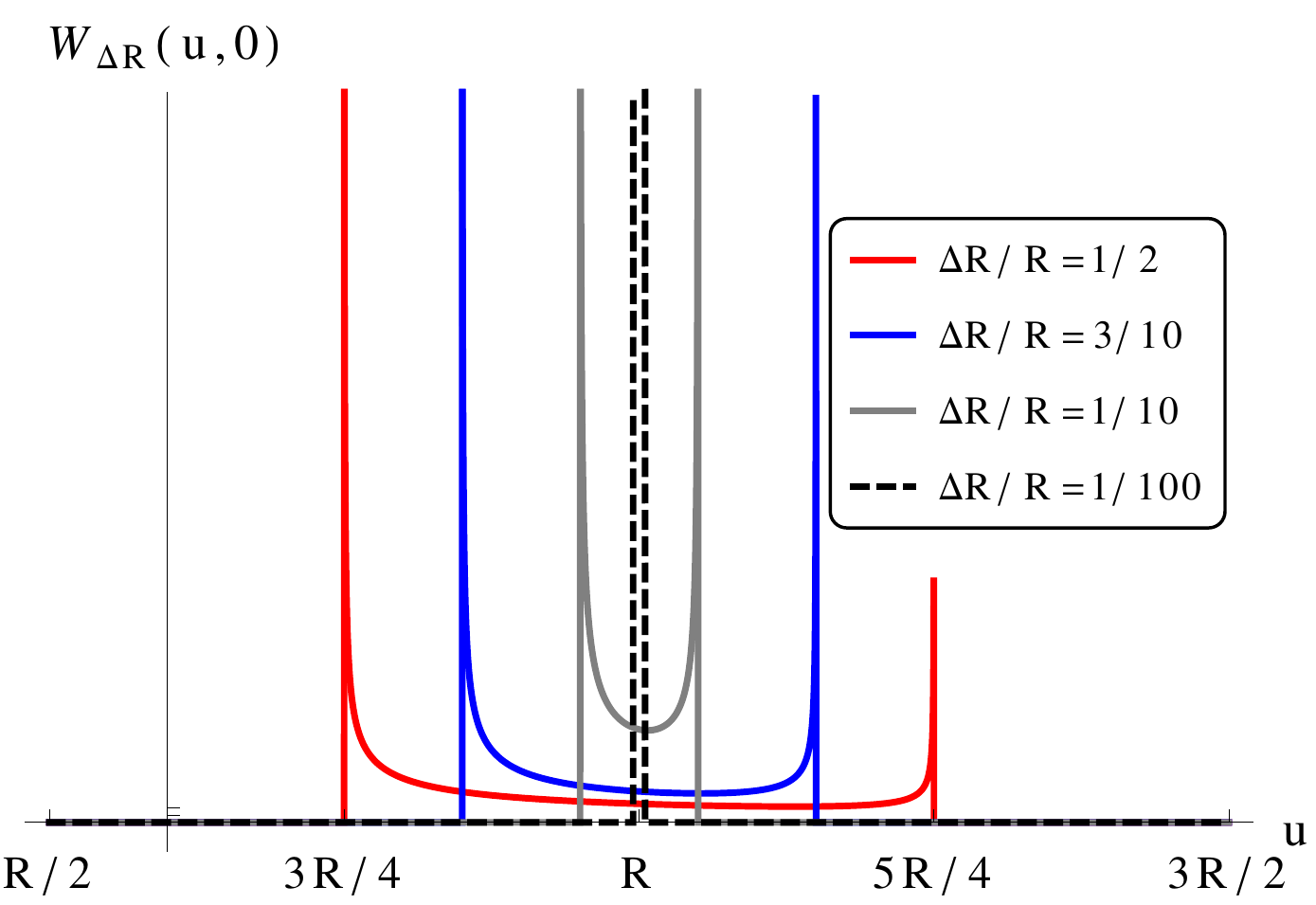}
    \caption[A weight function for the circular SE.]{A weight function for the circular SE. (left) $\alpha_{\nu}$ as a function of the thickness. (center) Scalar electric potential in the gap. (right) Weight function. }
\label{alphaCircularSEFig}
\end{figure}

Since $\Phi_{\mathcal{G}N}$ is set to zero, then

\[
\boldsymbol{\Theta}(\boldsymbol{r}) = \frac{1}{2\pi}\lim_{N\rightarrow\infty} \sum_{m=1}^{N-1} (\Phi_{\mathcal{G}m}-\Phi_{\mathcal{G}m+1})  \rcirclerightint_{\partial \mathcal{G}_m^{+}(0)} \frac{d\boldsymbol{r}'}{|\boldsymbol{r}-\boldsymbol{r}'|}    
\]

using a Taylor series expansion of the scalar potential $\Phi_{\mathcal{G}m+1} = \Phi_{\mathcal{G}m} + \delta u_m \partial_u \Phi_{\mathcal{G}}(u_m)$ then

\[
\boldsymbol{\Theta}(\boldsymbol{r}) = -\frac{1}{2\pi}\lim_{N\rightarrow\infty} \sum_{m=1}^{N-1} \left. \frac{\partial \Phi_{\mathcal{G}}}{\partial u} \right|_{u=u_m} \delta u_m  \rcirclerightint_{\partial \mathcal{G}_m^{+}(0)} \frac{d\boldsymbol{r}'}{|\boldsymbol{r}-\boldsymbol{r}'|} = -\frac{1}{2\pi} \int_{R_{-}}^{R_{+}} \frac{\partial \Phi_{\mathcal{G}}}{\partial u''} du'' \rcirclerightint_{\partial \mathcal{G}(u'')} \frac{d\boldsymbol{r}'}{|\boldsymbol{r}-\boldsymbol{r}'|}    
\]

which can be written as follows

\begin{equation}
\boxed{
\boldsymbol{\Theta}(\boldsymbol{r}) = \left\langle \Theta^{\mbox{\tiny \textbf{GSE}}}(\boldsymbol{r};u'') \right\rangle_{\mathcal{G}} = \int_{R_{-}}^{R_{+}} \mathscr{W}_{\Delta R}(u'') \Theta^{\mbox{\tiny \textbf{GSE}}}(\boldsymbol{r};u'') du''
}\hspace{0.25cm}\mbox{Electric vector potential of the gapped circular SE}
\label{electricVectorPotentialGappedCircularSEEq}
\end{equation}

where
\[
\Theta^{\mbox{\tiny \textbf{GSE}}}(\boldsymbol{r};u'') = \frac{V_o}{2\pi} \rcirclerightint_{\partial \mathcal{G}(u'')} \frac{d\boldsymbol{r}'}{|\boldsymbol{r}-\boldsymbol{r}'|}
\]
is the gapless solution with $\mathcal{G}(u'')$ a circle of radius $u''$ and
\begin{equation}
\mathscr{W}_{\Delta R}(u) = -\frac{1}{V_o} \frac{\partial \Phi_{\mathcal{G}}}{\partial u}
\label{weightFunctionCircularSEEq}
\end{equation}
plays the role of a weight function depending on the gap thickness $\nu=\Delta R$. In general, the GSE is a system that cannot be carried out experimentally in electrostatic conditions because in this setup two conductive sheets at different potential are put in contact. Thus the GSE is often used as a theoretical approximation of the real systems that include gaps of small thickness. However, in this section is shown that GSE can be also employed to study the gapped case. This is just the result in Eq.~(\ref{electricVectorPotentialGappedCircularSEEq}) where electric vector potential of the gapped SE at any point in the $\mathbb{R}^3$ space can be found by taking weighted average of the gapless solution on the gap. 

\begin{figure}[H]
\centering %system.pdf
\includegraphics[width=0.33\textwidth]{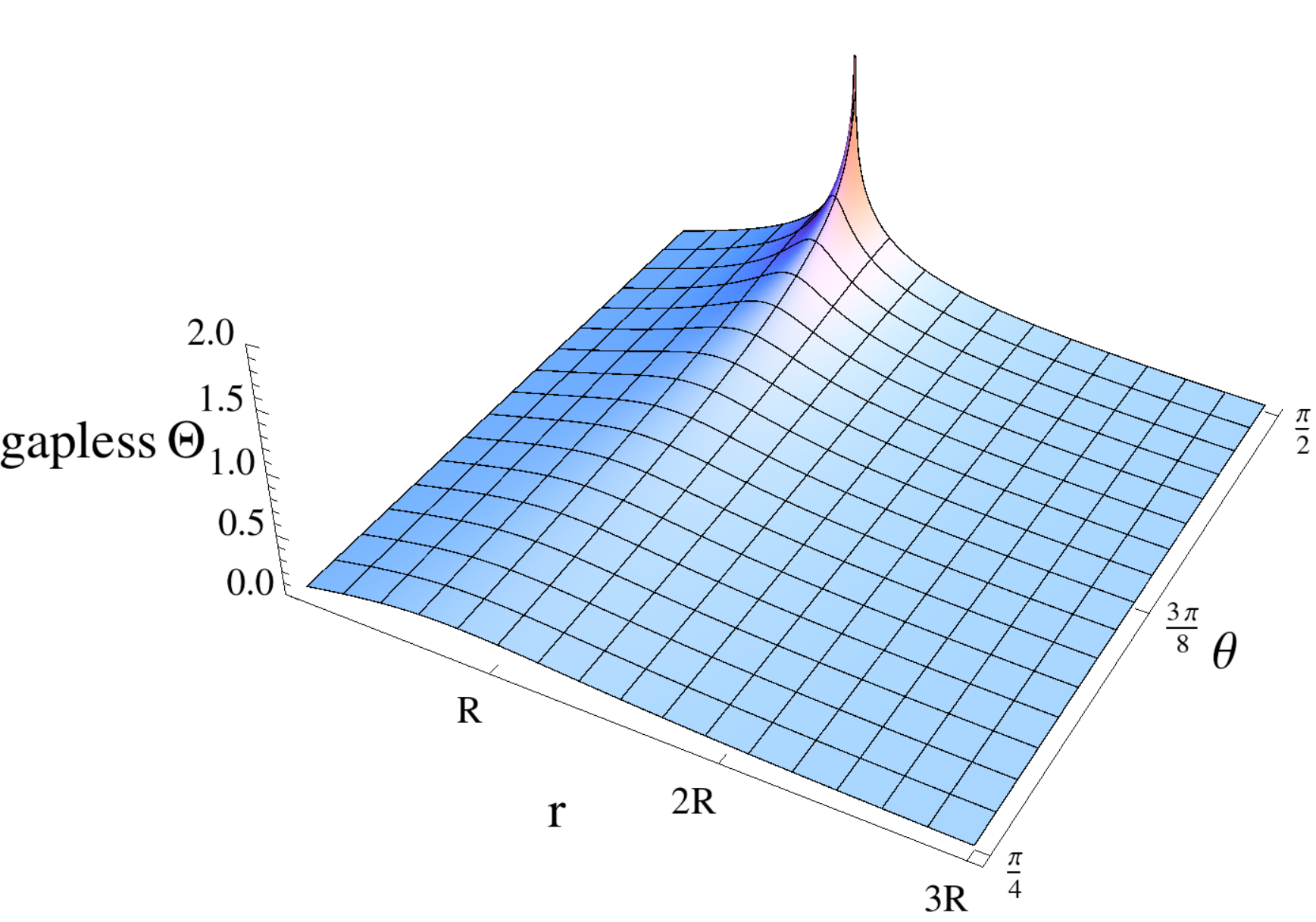}
\includegraphics[width=0.33\textwidth]{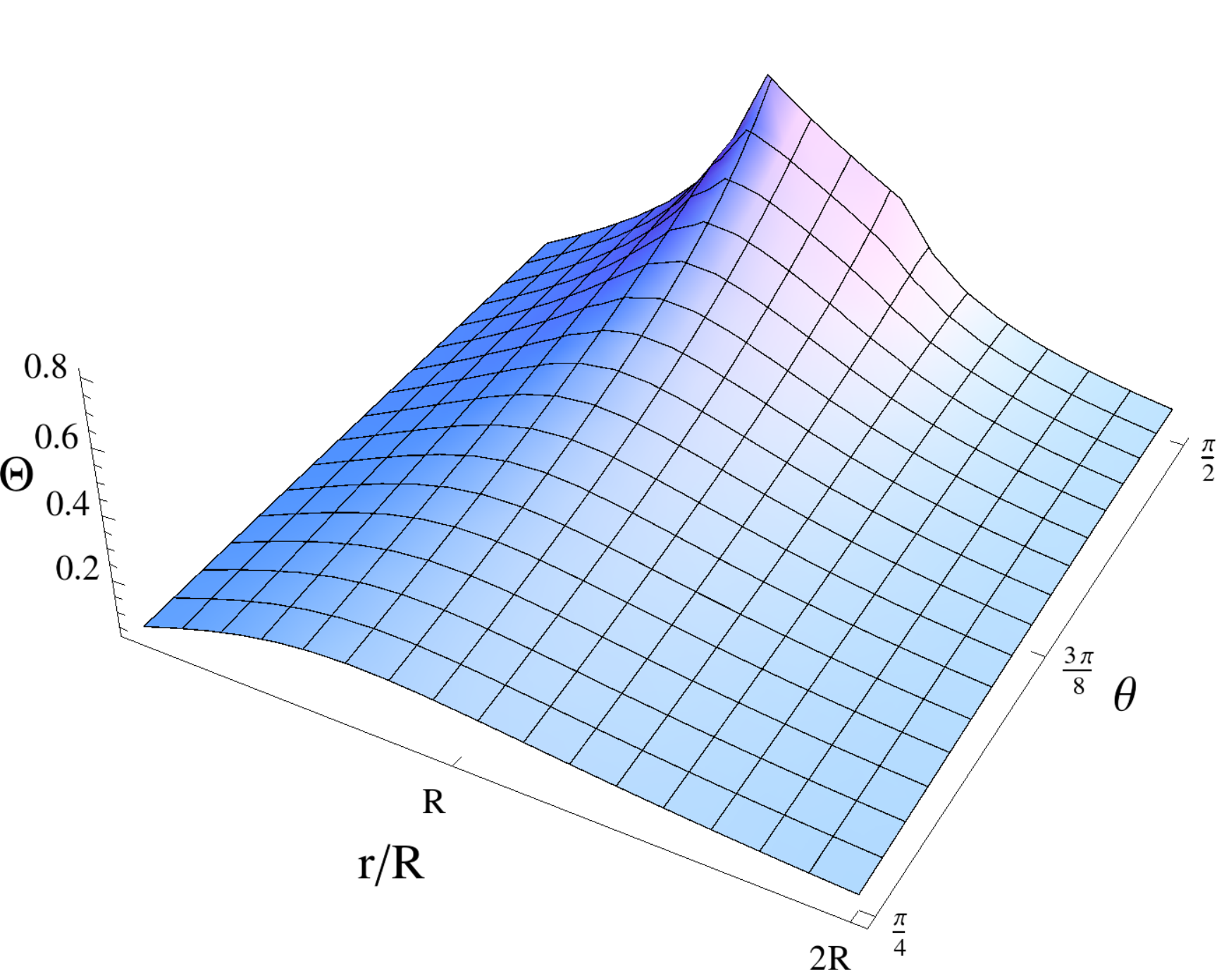}
\includegraphics[width=0.32\textwidth]{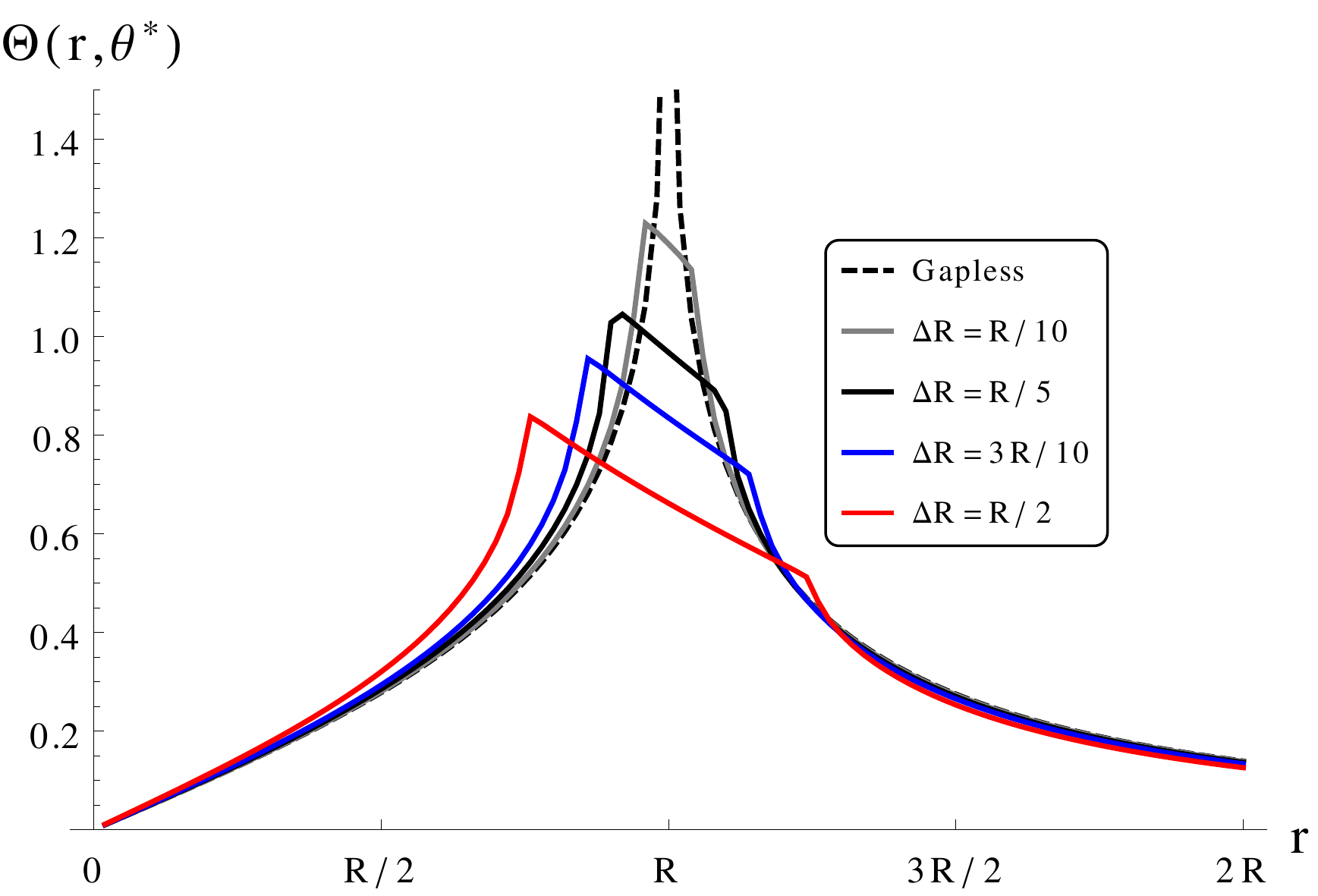}
    \caption[Electric Vector Potential of the Circular SE.]{Electric Vector Potential of the Circular SE. (left) Gapless circula SE, (center) Gapped circular SE with $\nu = \Delta R= R/2$. (right) Electric Vector Potential at $\theta = 0.999 \pi/2$ and different thickness values.}
\label{gappedPHICircularlarSEFig}
\end{figure}

Other quantities of interest as the electric field and surface charge density can be found from Eq.~(\ref{electricVectorPotentialGappedCircularSEEq}) straightforwardly since differential operators as the curl are applied on the unprimed coordinates 
\[
\boldsymbol{E}(\boldsymbol{r}) = \nabla\times\left\langle \boldsymbol{\Theta}^{\mbox{\tiny \textbf{GSE}}}(\boldsymbol{r};u'') \right\rangle_{\mathcal{G}} = \left\langle \nabla\times\boldsymbol{\Theta}^{\mbox{\tiny \textbf{GSE}}}(\boldsymbol{r};u'') \right\rangle_{\mathcal{G}} = \left\langle \boldsymbol{E}^{\mbox{\tiny \textbf{GSE}}}(\boldsymbol{r};u'') \right\rangle_{\mathcal{G}}
\]
therefore
\begin{equation}
\boldsymbol{E}(\boldsymbol{r}) = \left\langle \boldsymbol{E}^{\mbox{\tiny \textbf{GSE}}}(\boldsymbol{r};u'') \right\rangle_{\mathcal{G}} = \int_{R_{-}}^{R_{+}} \mathscr{W}_{\Delta R}(u'') \boldsymbol{E}^{\mbox{\tiny \textbf{GSE}}}(\boldsymbol{r};u'') du''
\hspace{0.25cm}\mbox{Electric field}
\label{electricFieldGappedCircularSEEq}
\end{equation}
and
\begin{equation}
\sigma(\boldsymbol{r}) = \left\langle \sigma^{\mbox{\tiny \textbf{GSE}}}(\boldsymbol{r};u'') \right\rangle_{\mathcal{G}} = \int_{R_{-}}^{R_{+}} \mathscr{W}_{\Delta R}(u'') \sigma^{\mbox{\tiny \textbf{GSE}}}(\boldsymbol{r};u'') du''
\hspace{0.25cm}\mbox{Surface charge density}
\label{electricSurfaceChargeDensityGappedCircularSEEq}
\end{equation}
are also weighted averages of their corresponding gapless solutions on the gap. 
\subsection{An approximation for the weight function}
According to the boundary conditions Eqs.~(\ref{DirichletBoundaryConditionsEq}) and (\ref{NeumannBoundaryConditionsEq}) the scalar electric potential in the gap region $\mathcal{G}$ is unknown as well as its derivative with respect the radial coordinate on the plane. In general, the potential in the gap depends on the several variables as the curvature of $\partial \mathscr{A}$ and $\nu$. However, as author of Ref.~\cite{schmied2010electrostatics} noted, the electric scalar potential on the gap of SE with curve contours can be approximated from the analytic expressions found for an Infinite Straight Gap SE. The scalar electric potential of the circular SE on the plane $z=0$ can be written as follows  
\begin{equation}
\Phi(x,y,0) = \Phi_{\Delta R}(u) =  V_o [1-H(u-R_{-})] +  \Phi_{SL}(u) + \alpha_{\Delta R} \xi_{\Delta R}(u) 
    \label{PhiOnThePlaneApproxEq}
\end{equation}
where $\alpha_{\Delta R}$ is a parameter depending on the gap (and eventually on the contour shape), $H(u)$ is the Heavy side step function
\[
H(z) = 0 \hspace{0.25cm}\textbf{if}\hspace{0.25cm}  z < 1 \hspace{0.25cm}\textbf{otherwise}\hspace{0.25cm} 1  ,
\]
the potential in the gap is given by
\[
\Phi_{SL}(u) = V_o \Pi_{\Delta R}(u) \left[ \frac{1}{2} - \frac{1}{\pi}\arcsin\left(\frac{2(u-R)}{\Delta R}\right)  \right] \hspace{0.25cm},\hspace{0.25cm}\xi_{\Delta R}(u) = V_o \Pi_{\Delta R}(u) \sqrt{1-\left[\frac{2}{\Delta R}(u-R)\right]^2}
\]
where $\Phi_{SL}(u)$ is the scalar potential of a Infinite Straight Gap SE of thickness $\nu =\Delta R$ or gap interpolation potential, $\xi_{\Delta R}(u)$ is a gap polarization potential and 
\[
\Pi_{\Delta R}(u) = 1 \hspace{0.25cm}\textbf{if}\hspace{0.25cm}  R_{-}<u<R_{+} \hspace{0.25cm}\textbf{otherwise}\hspace{0.25cm} 0
\]
is a boxcar function. Thus, the weight function (see Eq.~(\ref{weightFunctionCircularSEEq})) takes the form
\begin{equation}
\mathscr{W}_{\Delta R}(u)  =  \left(\frac{2}{\pi \Delta R} + \alpha_{\Delta R} \frac{4(u-R)}{\Delta R^2} \right) \frac{\Pi_{\Delta R}(u)}{\sqrt{1-\left[\frac{2}{\Delta R}(u-R)\right]^2}}.
\label{WFunctionCircularSEApproxEq}
\end{equation}

\begin{figure}[H]
\centering %system.pdf
\includegraphics[width=0.45\textwidth]{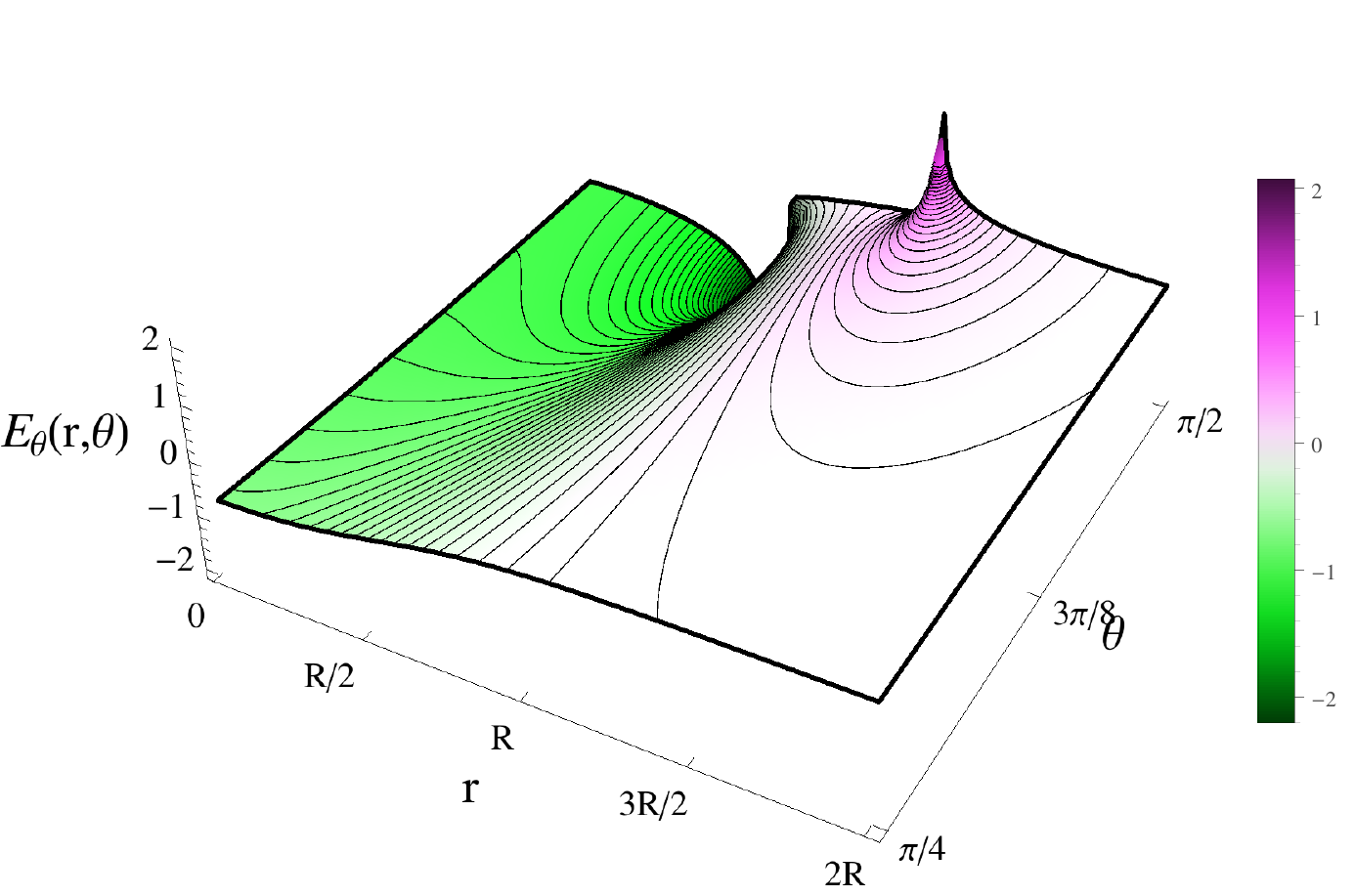}\hspace{0.5cm}
\includegraphics[width=0.45\textwidth]{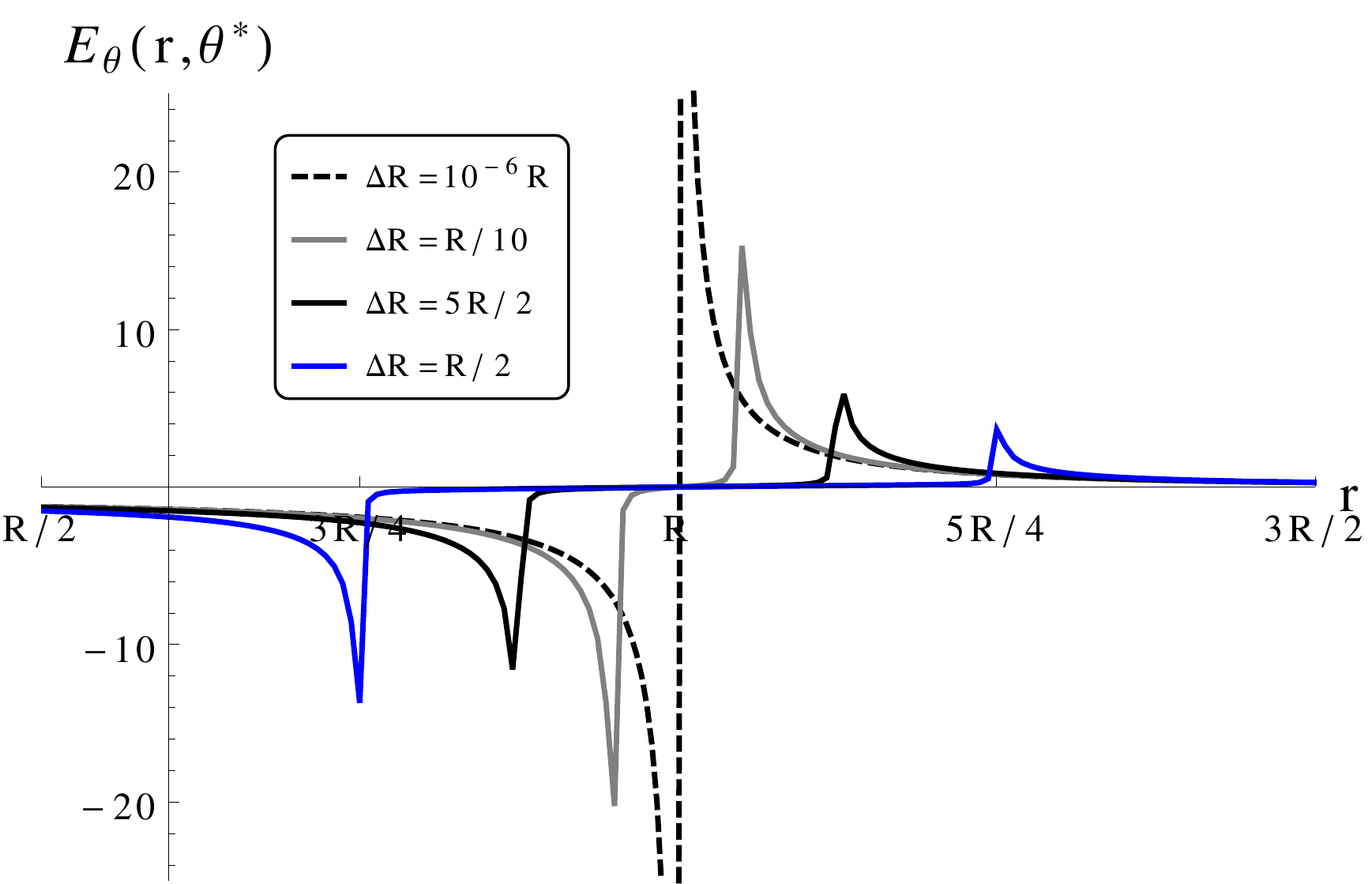}
    \caption[Electric Vector of the gapped circular SE on the Plane]{Electric Vector Potential of the Circular SE. (left) $\theta$-component of the electric field of  gapped circular SE with thickness $\nu=\Delta R=R/2$ in the $\mathbb{R}^3$ space. (right) $\boldsymbol{E}(r,\theta)$ evaluated at $\theta=\theta^*=\pi/2$ (near the plane $z=0$) for different values of the gap thickness.}
\label{EThetaOnPlaneCircularlarSEFig}
\end{figure}

The idea behind the technique, is to adjust the value of the parameter $\alpha_{\Delta R}$ by using the condition of null charge density in the gap mid point. The surface charge density is 
\[
\sigma(x,y) = 2\epsilon_o \lim_{z'\rightarrow 0^{+}} E_z(\boldsymbol{r}) = -\frac{\epsilon_o}{\pi}  \int_{\mathbb{R}^2}  \lim_{z'\rightarrow 0^{+}} \frac{\partial}{\partial z}\left( \frac{z-z'}{|\boldsymbol{r}-\boldsymbol{r}'|^3} \right) \Phi(x',y',0)  d^2 \boldsymbol{r}'  
\]
replacing Eq.~(\ref{PhiOnThePlaneApproxEq}) in the previous equation, then
\[
\sigma(x,y) = \sigma^{\mbox{\tiny \textbf{GSE}}}(x,y;R_{-}) - \lim_{z'\rightarrow 0^{+}} \left\{J[\Phi_{SL}](\boldsymbol{r}) + \alpha_{\Delta R} J[\xi_{\Delta R}](\boldsymbol{r}) \right\}
\]
where $J[f](\boldsymbol{r})$ is defined as follows
\[
J[f](\boldsymbol{r}) = \int_{\mathscr{G}}   \frac{(x-x')^2+(y-y')^2-2z^2}{[(x-x')^2+(y-y')^2+z^2]^{5/2}} f(x',y')  d^2 \boldsymbol{r}'
\]
or 
\[
J[f](r,\theta,\phi)=\int_{0}^{2\pi}d\phi'\int_{R_{-}}^{R_{+}} \frac{r^2(1-3\cos^2\theta)+u'^2-2ru'\cos(\phi-\phi')\sin\theta}{[r^2+u'^2-2ru'\cos(\phi-\phi')\sin\theta]^{5/2}} u'du' f(u')
\]
if $f$ is a radial function. 

\begin{figure}[H]
\centering %system.pdf
\includegraphics[width=0.6\textwidth]{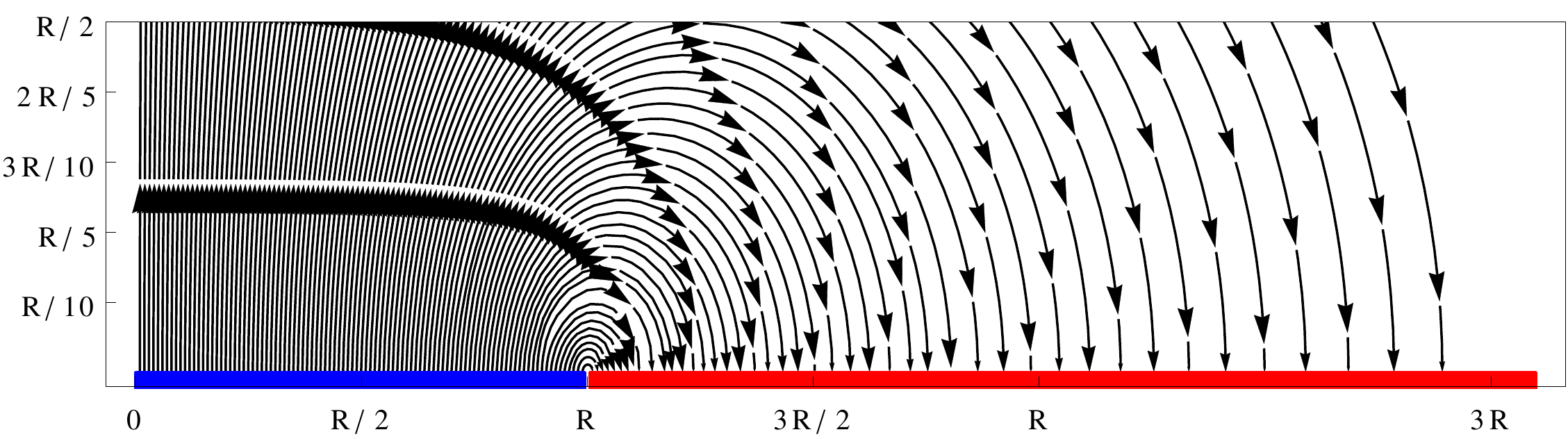}\\
\includegraphics[width=0.6\textwidth]{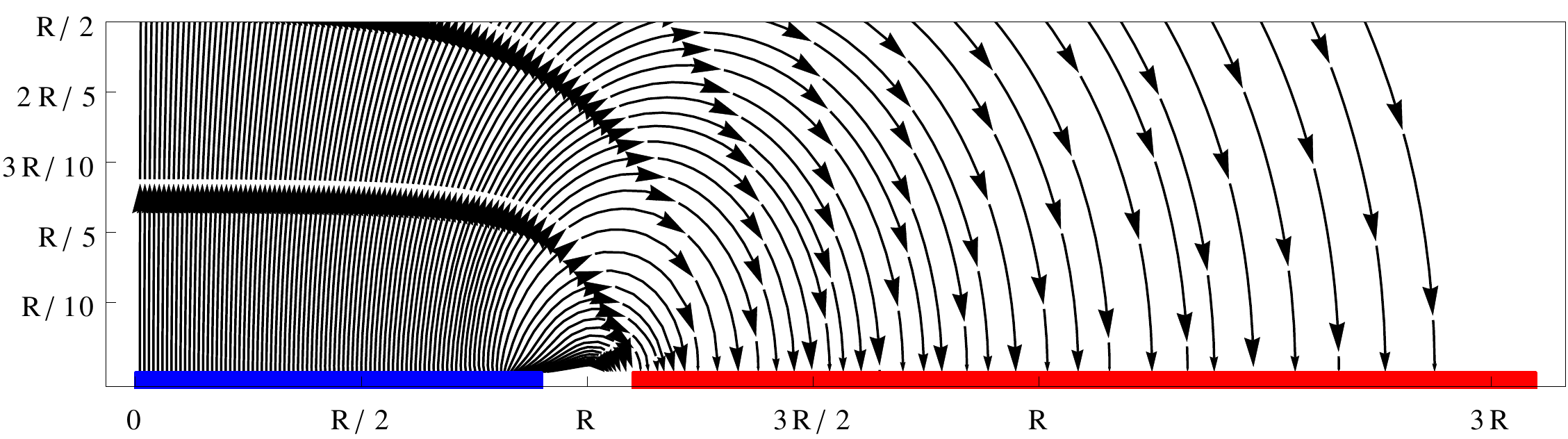}\\
\includegraphics[width=0.6\textwidth]{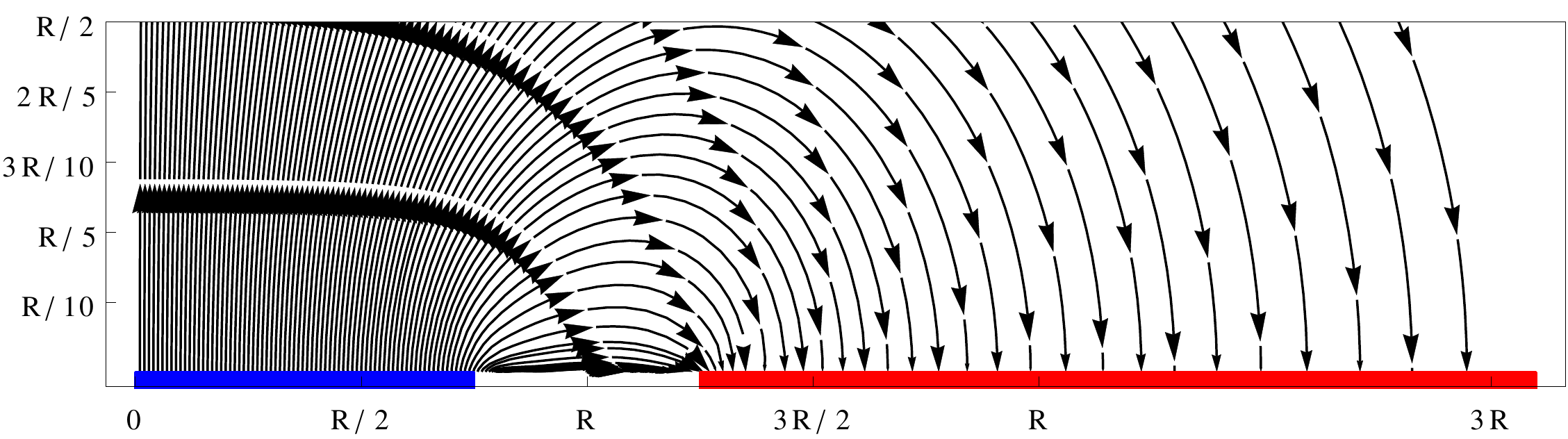}
    \caption[Vector field.]{Vector Field. Top to bottom. Cross section of the electric vector field for the circular gapped SE with $\nu=\Delta R=R/100, R/5$ and $R/2$. The scalar electric potential of the blue and red sheets are $Vo$ and zero respectively. }
\label{streamLinesCircularSEFig}
\end{figure}

In order to find $\alpha_{\Delta R}$, it is used the condition of null charge density in the mid trajectory $\partial \mathcal{A}$, this is $\sigma(x,y)=0$ at $(r=R,\theta=\pi/2,\phi)$. The result is 
\begin{equation}
\alpha_{\Delta R}(\phi) = - \frac{\frac{\pi}{\epsilon_o}\sigma^{\mbox{\tiny \textbf{GSE}}}(R\cos(\phi),R\sin\phi;R_{-})+J[f](R,\pi/2,\phi)}{J[f](R,\pi/2,\phi)}.
\label{alphaApproxEq}    
\end{equation}
For the case of the circular SE, $\sigma^{\mbox{\tiny \textbf{GSE}}}(x,y;R_{-})$ and $J[f](r,\theta,\phi)$ are independent of the $\phi$-coordinate and $\phi$ can be set as zero in the previous formula to compute $\alpha_{\Delta R}$. The result of $\alpha_{\Delta R}$ as a function of the thickness is shown in Fig.~(\ref{alphaCircularSEFig})-left by using a standard trapezoidal method to evaluate the integrals. This parameter vanishes as the thickness of the gap decreases $\lim_{\Delta R \to 0}\alpha_{\Delta R} = 0$. A plot of the scalar electric potential on the plane and the weight function given by Eqs.~(\ref{PhiOnThePlaneApproxEq}) and (\ref{WFunctionCircularSEApproxEq}) are also shown in Fig.~(\ref{alphaCircularSEFig}). The electric vector potential of the gapped circular SE can be computed by evaluating numerically the average in Eq.~(\ref{electricVectorPotentialGappedCircularSEEq}) using the gapless solution in Eq.~(\ref{ThetaPhiGaplessEq}) and the weight function in Eq.~(\ref{WFunctionCircularSEApproxEq}). A plot of $\boldsymbol{\Theta}$ for the gapped circular SE is shown in Fig.~(\ref{gappedPHICircularlarSEFig}). The electric vector potential of the gapless circular SE is divergent at $(r,\theta)=(R,\pi/2)$ since the elliptic integral of the firs kind $K(z)$ in Eq.~(\ref{ThetaPhiGaplessEq}) diverges, $\lim_{z \to 1} K(z) \rightarrow \infty$. Once the gap emerges on the SE, the vector electric potential becomes finite at any point of the space including the $xy$-plane (see Fig.~(\ref{gappedPHICircularlarSEFig})-right), and this feature of $\Theta(\boldsymbol{r})$ is inherited by the electric field and the surface charge density when the gap is considered. Similarly, the electric field of the gapped circular SE is computed from Eq.~(\ref{electricFieldGappedCircularSEEq}) using the gapless solutions given by Eqs.~(\ref{ErGaplessEq}) and (\ref{ThetaPhiGaplessEq}) and the weight function in Eq.~(\ref{WFunctionCircularSEApproxEq}). A plot of the $\theta$-component of the electric field is shown in Fig.~\ref{EThetaOnPlaneCircularlarSEFig}. In general, the Neumman condition of the scalar electric potential \ref{NeumannBoundaryConditionsEq} implies that the $z$-component of the electric field on the plane $E_z(x,y,0)=-\lim_{\theta\to\pi/2}E_{\theta}(r,\theta)$ must be zero on the gap region $\mathscr{G}$. It should be noted that is approximately fulfilled by $E_{\theta}(r,\theta)$ as it is shown in Fig.~\ref{EThetaOnPlaneCircularlarSEFig}-right where $E_{\theta}(r,\pi/2)$ is approximately zero in $\mathscr{G}$ except near the boundaries $\partial\mathscr{A}_{-}$ and $\partial\mathscr{A}_{+}$. This discrepancy near the SE boundary occurs because the computation of $\alpha_{\Delta R}$ via Eq.~(\ref{alphaApproxEq}) only ensures that charge density is exactly zero at the the mid trajectory $\partial \mathscr{A}$. 
\subsection{Gapped SE of generic contour}
According to Eq.~(\ref{OmegaUsefulEq})
\[
\boldsymbol{\Omega}(\boldsymbol{r}) = \frac{1}{4\pi} \int_{\mathbb{R}^2} \frac{\hat{z}\times\boldsymbol{E}(\boldsymbol{r}')}{|\boldsymbol{r}-\boldsymbol{r}'|} d^2 \boldsymbol{r}' = \frac{1}{4\pi} \int_{\mathbb{R}^2 / \mathcal{A}_{-}\cup\mathcal{A}_{+}} \frac{u'du'd\phi'}{|\boldsymbol{r}-\boldsymbol{r}'|} \left[E_u(\boldsymbol{r}')\hat{\phi}(\phi')-E_\phi(\boldsymbol{r}')\hat{u}(\phi')\right] 
\]
The integration can be performed in $\mathbb{R}^2 / \mathcal{A}_{-}\cup\mathcal{A}_{+}$ rather than the whole $xy$-plane since $\lim_{z \to 0} \boldsymbol{E}(\boldsymbol{r})$ must be perpendicular to the $z=0$ plane in the regions $\mathcal{A}_{-}$ and $\mathcal{A}_{+}$. This feature can be seen in the the electric field streamlines in Fig.~\ref{streamLinesCircularSEFig} for the gapped circular SE. It occurs due to conductor sheets spread in regions $\mathcal{A}_{-}$ and $\mathcal{A}_{+}$ must be equipotential surfaces of $\Phi(\boldsymbol{r})$. The previous equation can be written more explicitly as follows
\[
\boldsymbol{\Omega}(\boldsymbol{r}) = \frac{1}{4\pi} \int_{\mathscr{G}} \frac{d^2 \boldsymbol{r}'}{|\boldsymbol{r}-\boldsymbol{r}'|} \lim_{z' \to 0}\left[-\frac{\partial \Phi}{\partial u'}(\boldsymbol{r}')\hat{\phi}(\phi')+\frac{1}{u'}\frac{\partial \Phi}{\partial \phi'}(\boldsymbol{r}')\hat{u}(\phi')\right]
\]
therefore, the electric potential of the gapped SE takes the form
\begin{equation}
\boldsymbol{\Theta}(\boldsymbol{r}) = \frac{V_o}{2\pi} \int_{\mathscr{G}} \frac{d^2 \boldsymbol{r}'}{|\boldsymbol{r}-\boldsymbol{r}'|} \boldsymbol{\mathscr{W}}_{\nu}(\boldsymbol{r}')
\label{ThetaGenericGappedSEFirstEq}
\end{equation}
where $\boldsymbol{\mathscr{W}}_{\nu}(u,\phi)$ is a vector depending on the gap thickness $\nu$ with units of inverse length whose components components are
\begin{equation}
(\boldsymbol{\mathscr{W}}_{\nu})_u = \frac{1}{V_o} \frac{1}{u}\frac{\partial \Phi_{\mathscr{G}}}{\partial \phi}(u,\phi) \hspace{0.5cm}\mbox{and}\hspace{0.5cm}(\boldsymbol{\mathscr{W}}_{\nu})_\phi = -\frac{1}{V_o} \frac{\partial \Phi_{\mathscr{G}}}{\partial u}(u,\phi)
\label{weightVectorEq}
\end{equation}
we use the same notation of the weight function of the circular SE given by Eq.~(\ref{weightFunctionCircularSEEq}). The Eqs.~(\ref{electricVectorPotentiaEq}) and (\ref{ThetaGenericGappedSEFirstEq}) are equivalent, and the gapples solution $\boldsymbol{\Theta}^{\mbox{\tiny \textbf{GSE}}}(\boldsymbol{r})$ given by Eq.~(\ref{electricVectorPotentiaGaplessEq}) is an implicit result that should be contained implicitly in Eq.~(\ref{ThetaGenericGappedSEFirstEq}). It is possible to proof this by considering the gapless condition on the scalar potential
\[
\lim_{\nu \to 0} \Phi_{\mathscr{G}}(u,\phi) = V_o[1-H(|(u,\phi)-(\mathscr{R}(\phi),\phi)|)]
\]
with $(\mathscr{R}(\phi),\phi)$ the parametric representation of an arbitrary closed contour $\partial A$ and $H(z)$ the heavyside step function. This limit condition can be also observed in Fig.~\ref{alphaCircularSEFig}-center for the particular case $\mathscr{R}(\phi)=R$. The components of the weight vector $\boldsymbol{\mathscr{W}}_{\nu}(\boldsymbol{r})$ in the gapless limit are
\[
\lim_{\nu \to 0} (\boldsymbol{\mathscr{W}}_{\nu})_u =  \frac{1}{V_o} \frac{1}{u} \left[ - V_o \frac{\partial}{\partial\phi} H(u-\mathscr{R}(\phi))  \right] = \frac{\dot{\mathscr{R}}(\phi)}{u} \delta(u-\mathscr{R}(\phi)) 
\]
and
\[
\lim_{\nu \to 0} (\boldsymbol{\mathscr{W}}_{\phi})_u =  -\frac{1}{V_o} \left[ - V_o \frac{\partial}{\partial u} H(u-\mathscr{R}(\phi))  \right] = \delta(u-\mathscr{R}(\phi)) 
\]
with $\delta(z)$ the dirac delta function, hence
\begin{equation}
\boldsymbol{\mathscr{W}}_{\nu \rightarrow 0}(\boldsymbol{r}) =  \delta(u-\mathscr{R}(\phi)) \left[\frac{\dot{\mathscr{R}}(\phi)}{u} \hat{u}(\phi) + \hat{\phi}(\phi)\right].
\label{weightVectorLimitEq}
\end{equation}
The Eq.~(\ref{ThetaGenericGappedSEFirstEq}) in the gapless limit becomes
\[
\lim_{\nu \to 0} \boldsymbol{\Theta}(\boldsymbol{r}) = \frac{V_o}{2\pi} \int_{\mathscr{G}} \frac{u'du'd\phi'}{|\boldsymbol{r}-\boldsymbol{r}'|} \delta(u'-\mathscr{R}(\phi')) \left[\frac{\dot{\mathscr{R}}(\phi')}{u'} \hat{u}(\phi') + \hat{\phi}(\phi')\right]
\]
evaluating the integral over $u'$
\[
\lim_{\nu \to 0} \boldsymbol{\Theta}(\boldsymbol{r}) = \frac{V_o}{2\pi} \int_{0}^{2\pi} \frac{\mathscr{R}(\phi') d\phi'}{|\boldsymbol{r}-(\mathscr{R}(\phi'),\phi')|} \left[\frac{\dot{\mathscr{R}}(\phi')}{\mathscr{R}(\phi')} \hat{u}(\phi') + \hat{\phi}(\phi')\right] = \frac{V_o}{2\pi} \int_{0}^{2\pi} \frac{\left[\dot{\mathscr{R}}(\phi') \hat{u}(\phi') + \mathscr{R}(\phi')\hat{\phi}(\phi')\right] d\phi'}{|\boldsymbol{r}-(\mathscr{R}(\phi'),\phi')|} 
\]
this is
\[
\lim_{\nu \to 0} \boldsymbol{\Theta}(\boldsymbol{r}) = \frac{V_o}{2\pi} \rcirclerightint_{\partial \mathcal{A}} \frac{d\boldsymbol{r}'}{|\boldsymbol{r}-\boldsymbol{r}'|} = \boldsymbol{\Theta}^{\mbox{\tiny \textbf{GSE}}}(\boldsymbol{r})
\]
since $d\boldsymbol{r}'=\left[\dot{\mathscr{R}}(\phi') \hat{u}(\phi') + \mathscr{R}(\phi')\hat{\phi}(\phi')\right] d\phi'$ recovering Eq.~(\ref{electricVectorPotentiaGaplessEq}) as we expected. 

\subsection{The magnetic analogue of the gapped SE}
The electric field of the gapped SE can be obtained by applying the curl to Eq.~(\ref{ThetaGenericGappedSEFirstEq}). To this aim, we may use 
\[
\nabla \times \frac{\boldsymbol{\mathscr{W}}_{\nu}(\boldsymbol{r}')}{|\boldsymbol{r}-\boldsymbol{r}'|} = \nabla \left( \frac{1}{|\boldsymbol{r}-\boldsymbol{r}'|} \right) \times \boldsymbol{\mathscr{W}}_{\nu}(\boldsymbol{r}') = \boldsymbol{\mathscr{W}}_{\nu}(\boldsymbol{r}') \times \frac{\boldsymbol{r}-\boldsymbol{r}'}{|\boldsymbol{r}-\boldsymbol{r}'|^3}. 
\]
therefore
\begin{equation}
\boldsymbol{E}(\boldsymbol{r}) = \frac{V_o}{2\pi} \textbf{\mbox{sgn}}(z) \int_{\mathscr{G}}  \frac{\boldsymbol{\mathscr{W}}_{\nu}(\boldsymbol{r}')d^2 \boldsymbol{r}' \times (\boldsymbol{r}-\boldsymbol{r}')}{|\boldsymbol{r}-\boldsymbol{r}'|^3}    \hspace{0.25cm} \mbox{BSL law for the gapped SE} 
\label{bslEFieldGappedEq}
\end{equation}
is a generalization of the Eq.~(\ref{BSLElectricPontentialFromThetaEq}) but including a gap of thickness $\nu$ between both electrodes. Of course, we may recover the gapless result Eq.~(\ref{BSLElectricPontentialFromThetaEq}) from Eq.~(\ref{bslEFieldGappedEq}) by using Eq.~(\ref{weightVectorLimitEq}). Note that Eq.~(\ref{bslEFieldGappedEq}) is similar to the one employed for the computation of the magnetic field $\boldsymbol{B}(\boldsymbol{r})$ due to a ribbon on a circuit $\mathscr{G}$ of thickness $\nu$ carrying a current surface density $\boldsymbol{K}$
\[
\boldsymbol{B}(\boldsymbol{r}) = \frac{\mu_o}{4\pi} \int_{\mathscr{G}}  \frac{\boldsymbol{K}(\boldsymbol{r}')d^2 \boldsymbol{r}' \times (\boldsymbol{r}-\boldsymbol{r}')}{|\boldsymbol{r}-\boldsymbol{r}'|^3}
\]
which is a formulation of the Biot-Savart law in magnetostatics. In the electrostatic problem, the weight vector $\boldsymbol{\mathscr{W}}_{\nu}(\boldsymbol{r})$ is playing the role of density current $\boldsymbol{K}(\boldsymbol{r})$. Sign function in Eq.~(\ref{bslEFieldGappedEq}) is absent in the biot-savart magnetic analogue since the electric field is not a fully solenoidal field as occurs with the magnetic field.  The divergence in polar coordinates of the weight vector is
\[
\nabla \cdot \boldsymbol{\mathscr{W}}_{\nu}(\boldsymbol{r}) = \frac{1}{u}\frac{\partial}{\partial u}\left[u(\boldsymbol{\mathscr{W}}_{\nu})_u\right] + \frac{1}{u}\frac{\partial}{\partial \phi}\left[(\boldsymbol{\mathscr{W}}_{\nu})_{\phi}\right]
\]
where the components of the weight vector are given by Eq.~(\ref{weightVectorEq}), hence
\[
\nabla \cdot \boldsymbol{\mathscr{W}}_{\nu}(\boldsymbol{r}) = \frac{1}{u V_o}\left[ \frac{\partial^2 \Phi_{\mathcal{G}}}{\partial u\partial \phi} - \frac{\partial^2 \Phi_{\mathcal{G}}}{\partial \phi \partial u} \right] = 0.
\]
This is also a property that density current in magnetostatics has $\nabla \cdot \boldsymbol{K}(\boldsymbol{r}) = 0$. In the case of a bidimensional current ribbon loop, this continuity equation ensures the charge conservation, thus the incoming and outcoming charge of each point of the ribbon loop must be equal. Magnetostatics is described by using the Ampere's law     
\[
\nabla \times \boldsymbol{B}(\boldsymbol{r}) = \mu_o \boldsymbol{J}(\boldsymbol{r})
\]
with $\boldsymbol{J}(\boldsymbol{r})$ the volume charge density. For the case of a ribbon loop $\boldsymbol{J} = \delta(z)\boldsymbol{K}$ for each point belonging to the ribbon. At this point, one can formulate the following trick question : by virtue of the analogies and the Ampere's law, should $\nabla \times \boldsymbol{E}(\boldsymbol{r})$ be proportional to  $\boldsymbol{\mathscr{W}}_{\nu}(\boldsymbol{r})$ since the weight vector in the SE gap? Of course, We know that the answer to that question is no, since the $\nabla \times \boldsymbol{E}(\boldsymbol{r})$ is strictly zero in electrostatics. Even when, the weight vector $\boldsymbol{\mathscr{W}}_{\nu}(\boldsymbol{r})$ a non-zero vector field on the gap, the electric field given by Eq.~(\ref{bslEFieldGappedEq}) must be zero. To check that let us take the curl of this expression
\[
\nabla \times 
\boldsymbol{E}(\boldsymbol{r}) = \nabla \times \left\{ 2\textbf{\mbox{sgn}}(z) \frac{V_o}{4\pi}\int_{\mathbb{R}^3}  \frac{\delta(z)\boldsymbol{\mathscr{W}}_{\nu}(\boldsymbol{r}')d^3 \boldsymbol{r}' \times (\boldsymbol{r}-\boldsymbol{r}')}{|\boldsymbol{r}-\boldsymbol{r}'|^3} 
\right\}
\]
using the vector calculus identity Eq.~(\ref{vectorCalculusIdentityCurlEq}), then
\[
\nabla \times 
\boldsymbol{E}(\boldsymbol{r}) = 2\textbf{\mbox{sgn}}(z) \nabla \times \left\{  \frac{V_o}{4\pi}\int_{\mathbb{R}^3}  \frac{\delta(z)\boldsymbol{\mathscr{W}}_{\nu}(\boldsymbol{r}')d^3 \boldsymbol{r}' \times (\boldsymbol{r}-\boldsymbol{r}')}{|\boldsymbol{r}-\boldsymbol{r}'|^3} 
\right\} + 2\nabla \textbf{\mbox{sgn}}(z)\times\left\{\frac{V_o}{4\pi}\int_{\mathbb{R}^3}  \frac{\delta(z)\boldsymbol{\mathscr{W}}_{\nu}(\boldsymbol{r}')d^3 \boldsymbol{r}' \times (\boldsymbol{r}-\boldsymbol{r}')}{|\boldsymbol{r}-\boldsymbol{r}'|^3} 
\right\}
\]
this can be written as follows
\begin{equation}
    \nabla\times\boldsymbol{E}(\boldsymbol{r}) = 2\textbf{sgn}(z)\delta(z) V_o \left[  \boldsymbol{\mathscr{W}}_{\nu}(\boldsymbol{r}) + \hat{z} \times \frac{\boldsymbol{E}(\boldsymbol{r})}{V_o} \right]
    \label{curlEEq}
\end{equation}
if $z \neq 0$, where we have used $\nabla\textbf{sgn}(z) = 2\delta(z)\hat{z}$ and 
\[
\nabla \times \left\{  \frac{V_o}{4\pi}\int_{\mathbb{R}^3}  \frac{\delta(z)\boldsymbol{\mathscr{W}}_{\nu}(\boldsymbol{r}')d^3 \boldsymbol{r}' \times (\boldsymbol{r}-\boldsymbol{r}')}{|\boldsymbol{r}-\boldsymbol{r}'|^3} 
\right\} = V_o \boldsymbol{\mathscr{W}}_{\nu}(\boldsymbol{r}).
\]
Since $\delta(z)=0$ if $z \neq 0$, then we have $\nabla\times\boldsymbol{E}(\boldsymbol{r})=0$ from Eq.~(\ref{curlEEq}) for any point in the $\mathbb{R}^3$ not belonging to the SE and its gap as it should be.

\section*{Conclusion}
We demonstrated that it is possible to use standard vector calculus theorems to derive the electrostatic vector potential of the SE. Computation of electric field from the electric vector potential $\boldsymbol{\Theta}(\boldsymbol{r})$ is straightforward and algebraically less demanding than the one obtained from $\Phi(\boldsymbol{r})$ when no analogies with magnetostatics are used (see Section \ref{ElectricFieldSection}). As an illustrative example, the electric vector potential of circular SE, was used to determine its surface charge density finding that this system is globally neutral. As a future work, we expect that the strategy described in this document eventually could be used to map electrostatic problems to simpler ones including BSL integrals. One candidate is the non-planar SE where the magnetic analogue in general is unknown due to the surface curvature.

\section*{Acknowledgments}
This work was supported by Vicerrector\'ia de investigaci\'on, Universidad ECCI. Robert Salazar also thanks Fundaci\'on Colfuturo.
%----------------------------------------------------------------------------------------
%	REFERENCE LIST
%----------------------------------------------------------------------------------------
\bibliographystyle{ieeetr} %alpha, apalike, ieeetr
\bibliography{bibliography.bib}

\begin{appendices}

\section{Electric field from the electric scalar potential}
\label{ElectricFieldFromTheScalarPotentialSection}
The general solution for the Poisson's problem can written as
\[
\Phi(\boldsymbol{r}) = \frac{1}{4\pi\epsilon_o}\int_{\mathfrak{D}} \rho(\boldsymbol{r}) G(\boldsymbol{r},\boldsymbol{r}')d^3 \boldsymbol{r}' + \frac{1}{4\pi}\oint_{\partial \mathfrak{D}} \left[G(\boldsymbol{r},\boldsymbol{r}')\frac{\partial\Phi}{\partial n'} - \Phi(\boldsymbol{r'}) \frac{\partial G(\boldsymbol{r},\boldsymbol{r}')}{\partial n'} \right]d^2 \boldsymbol{r}'.
\]
Here the first term at the right hand vanishes since $\rho(\boldsymbol{r})=0$ for $r \in \mathfrak{D}$. This reduces to
\begin{equation}
\Phi(\boldsymbol{r}) = \frac{V_o}{2\pi} \int_{\mathcal{A}}  \lim_{z'\rightarrow 0^{+}} \frac{z-z'}{|\boldsymbol{r}-\boldsymbol{r}'|^3} dx'dy'
\label{PhiPlanarSEStandarEq}
\end{equation}
Then the electric field can be computed from
\[
\boldsymbol{E}(\boldsymbol{r}) = - \nabla \Phi(\boldsymbol{r}) = -\frac{V_o}{2\pi} \int_{\mathcal{A}} \nabla \left[ \lim_{z'\rightarrow 0^{+}} \frac{z-z'}{|\boldsymbol{r}-\boldsymbol{r}'|^3} \right] dx'dy'.
\]
In the previous expression the gradient is applied on the unprimed coordinates. We may use the following identity (see Eq.~(\ref{firstPropertyEq}) in Appendix \ref{AppendixUsefulIdentities})
\[
\nabla \left[ \lim_{z'\rightarrow 0^{+} } \frac{z-z'}{|\boldsymbol{r}-\boldsymbol{r}'|^3} \right] = - \left[ \lim_{z'\rightarrow 0^{+}} \nabla' \frac{z-z'}{|\boldsymbol{r}-\boldsymbol{r}'|^3} \right]
\]
where $\nabla'$ operates on the primed coordinates. Therefore
\[
\boldsymbol{E}(\boldsymbol{r}) = \frac{V_o}{2\pi} \int_{\mathcal{A}} \lim_{z'\rightarrow 0^{+} } \left\{ \frac{\partial}{\partial x'} \left[ \frac{(z-z')\hat{x}}{|\boldsymbol{r}-\boldsymbol{r}'|^3} \right] + \frac{\partial}{\partial y'} \left[ \frac{(z-z')\hat{y}}{|\boldsymbol{r}-\boldsymbol{r}'|^3} \right] + \frac{\partial}{\partial z'} \left[ \frac{(z-z')\hat{z}}{|\boldsymbol{r}-\boldsymbol{r}'|^3} \right] \right\} dx'dy'
\]
Again, we may use the Green's theorem Eq.~(\ref{greensTheoremEq}) identifying $M$ and $L$ as follows
\[
M \longrightarrow \frac{(z-z')\hat{x}}{|\boldsymbol{r}-\boldsymbol{r}'|^3} \hspace{0.5cm} \mbox{and} \hspace{0.5cm} L \longrightarrow -\frac{(z-z')\hat{y}}{|\boldsymbol{r}-\boldsymbol{r}'|^3}
\]
to write
\begin{equation}
\boldsymbol{E}(\boldsymbol{r}) = \frac{V_o}{2\pi} \oint_{\partial \mathcal{A}} \frac{-(z-z')\hat{y}dx'+(z-z')\hat{x}dy'}{|\boldsymbol{r}-\boldsymbol{r}'|^3} + \hat{z} E_z(\boldsymbol{r})
\label{EFieldAuxiliarEq}
\end{equation}
where the z-component of the electric field is 
\[
E_z(\boldsymbol{r}) = \frac{V_o}{2\pi} \int_{\mathcal{A}} \lim_{z'\rightarrow 0^{+} }  \frac{\partial}{\partial z'} \left[ \frac{(z-z')}{|\boldsymbol{r}-\boldsymbol{r}'|^3} \right] dx'dy' = - \frac{V_o}{2\pi} \int_{\mathcal{A}} \frac{\partial}{\partial z} \left[ \lim_{z'\rightarrow 0^{+} }   \frac{(z-z')}{|\boldsymbol{r}-\boldsymbol{r}'|^3} \right] dx'dy'.
\]
Now, we may use the following identity (see Subsection \ref{AppendixSecondIdentiySubsection} in the Appendix \ref{AppendixUsefulIdentities})
\[
\frac{\partial}{\partial z}\left[ \lim_{z' \to 0^{+}}\left(\frac{z-z'}{|\boldsymbol{r}-\boldsymbol{r}'|^3}\right) \right] = \frac{\partial}{\partial x'}\left[ \lim_{z' \to 0^{+}} \left(\frac{x-x'}{|\boldsymbol{r}-\boldsymbol{r}'|^3}\right) \right] + \frac{\partial}{\partial y'}\left[ \lim_{z' \to 0^{+}} \left(\frac{y-y'}{|\boldsymbol{r}-\boldsymbol{r}'|^3}\right) \right]
\]
in order to write 
\[
E_z(\boldsymbol{r}) =  - \frac{V_o}{2\pi} \int_{\mathcal{A}} \frac{\partial}{\partial x'}\left[ \lim_{z' \to 0^{+}} \left(\frac{x-x'}{|\boldsymbol{r}-\boldsymbol{r}'|^3}\right) \right] + \frac{\partial}{\partial y'}\left[ \lim_{z' \to 0^{+}} \left(\frac{y-y'}{|\boldsymbol{r}-\boldsymbol{r}'|^3}\right) \right] dx'dy'.
\]
This enables us to use the Green's theorem Eq.~(\ref{greensTheoremEq}) identifying $M$ and $L$ as follows
\[
M \longrightarrow \lim_{z' \to 0^{+}} \left(\frac{x-x'}{|\boldsymbol{r}-\boldsymbol{r}'|^3}\right) \hspace{0.5cm} \mbox{and} \hspace{0.5cm} L \longrightarrow -\lim_{z' \to 0^{+}} \left(\frac{y-y'}{|\boldsymbol{r}-\boldsymbol{r}'|^3}\right)
\]
then
\[
E_z(\boldsymbol{r}) =  - \frac{V_o}{2\pi} \int_{\mathcal{A}} \left[\frac{\partial}{\partial x'} M(x',y') - \frac{\partial}{\partial y'} L(x',y') \right]dx'dy' = - \frac{V_o}{2\pi}\oint_{\partial \mathcal{A}} L(x',y') dx' + M(x',y') dy'
\]
this is
\begin{equation}
E_z(\boldsymbol{r}) =  \frac{V_o}{2\pi}\oint_{\partial \mathcal{A}} \lim_{z' \to 0^{+}} \left(\frac{y-y'}{|\boldsymbol{r}-\boldsymbol{r}'|^3}\right) dx' - \lim_{z' \to 0^{+}} \left(\frac{x-x'}{|\boldsymbol{r}-\boldsymbol{r}'|^3}\right) dy' = \frac{V_o}{2\pi}\oint_{\partial \mathcal{A}} \lim_{z' \to 0^{+}} \frac{(y-y')dx'-(x-x')dy'}{|\boldsymbol{r}-\boldsymbol{r}'|^3}.
\label{EzIntegralEq}
\end{equation}
Replacing Eq.~(\ref{EzIntegralEq}) in Eq.~(\ref{EFieldAuxiliarEq}), it is obtained
\[
\boldsymbol{E}(\boldsymbol{r}) = - \nabla \Phi(\boldsymbol{r}) = \frac{V_o}{2\pi} \oint_{\partial \mathcal{A}} \frac{-(z-z')\hat{y}dx'+(z-z')\hat{x}dy' + [(y-y')dx'-(x-x')dy']\hat{z} }{|\boldsymbol{r}-\boldsymbol{r}'|^3} 
\]
Since
\[
d\boldsymbol{r}\times(\boldsymbol{r}-\boldsymbol{r}') =  \begin{vmatrix}
  \hat{x} & \hat{y} & \hat{z} \\
  dx' & dy' & 0 \\
  (x-x') & (y-y') & (z-z')
 \end{vmatrix} = \hat{x} dy'(z-z') - \hat{y} dy'(z-z') + \hat{z}[dx'(y-y') - dy'(x-x')] 
\]
the electric field takes the form
\[
\boxed{
\boldsymbol{E}(\boldsymbol{r}) = - \nabla \Phi(\boldsymbol{r}) = \frac{V_o}{2\pi} \rcirclerightint_{\partial \mathcal{A}}   \frac{d\boldsymbol{r}' \times (\boldsymbol{r}-\boldsymbol{r}')}{|\boldsymbol{r}-\boldsymbol{r}'|^3}
}
\]
for $z>0$, recovering the same result found with the electric vector potential (see Eq.~(\ref{BSLElectricPontentialFromThetaEq})) \footnote{It is interesting to see that electric field can be also found by noting that a SE defined by a path $\partial\mathcal{A}$ is the electrostatic analogue of a loop $\partial\mathcal{A}$ carrying an steady current as Authors of Ref.~\cite{oliveira2001biot} noted. In the free-current region the steady magnetic field is $\boldsymbol{B}=-\nabla \Psi$ where
\[
\Psi(\boldsymbol{r}) = \frac{\mu_o i_o}{4\pi} \int_{\mathcal{A}}  \lim_{z'\rightarrow 0^{+}} \frac{z-z'}{|\boldsymbol{r}-\boldsymbol{r}'|^3} dx'dy'
\]
is magnetic scalar potential. This formula is the magnetic analogue of the electric scalar potential given by Eq.~(\ref{PhiPlanarSEStandarEq}) where $\mu_o i_o/2$ is playing the role of $V_o$. The same identification can be obtained by comparing Eq.~(\ref{BSLElectricPontentialFromThetaEq}) and the standard Biot-Savart law 
\[
\boldsymbol{B}(\boldsymbol{r}) = \frac{\mu_o i_o}{4\pi} \oint_c  \frac{ d\boldsymbol{r}' \times (\boldsymbol{r}-\boldsymbol{r}') }{|\boldsymbol{r}-\boldsymbol{r}'|^3}.
\]
}.

\section{Useful identities}
\label{AppendixUsefulIdentities}

\subsection{First identity}
\label{AppendixFirstIdentiySubsection}
Using the product rule
\[
\nabla \left[f(\boldsymbol{r}) g(\boldsymbol{r})\right] = \left[\nabla f(\boldsymbol{r})\right] g(\boldsymbol{r}) + f(\boldsymbol{r}) \left[\nabla g(\boldsymbol{r})\right] 
\]
then
\[
\nabla \left[ \lim_{z'\rightarrow 0^{+} } \frac{z-z'}{|\boldsymbol{r}-\boldsymbol{r}'|^3} \right] = \lim_{z'\rightarrow 0^{+}} (z-z') \nabla \lim_{z'\rightarrow 0^{+} } \frac{1}{|\boldsymbol{r}-\boldsymbol{r}'|^3} + \left[\nabla \lim_{z'\rightarrow 0^{+}} (z-z')\right] \lim_{z'\rightarrow 0^{+} } \frac{1}{|\boldsymbol{r}-\boldsymbol{r}'|^3}
\]
Now
\[
\nabla \lim_{z'\rightarrow 0^{+} } \frac{1}{|\boldsymbol{r}-\boldsymbol{r}'|^3} = - \lim_{z'\rightarrow 0^{+} } \nabla'  \frac{1}{|\boldsymbol{r}-\boldsymbol{r}'|^3}  
\]
and
\[
\nabla \lim_{z'\rightarrow 0^{+}} (z-z') = \nabla z = \hat{z} = - \lim_{z'\rightarrow 0^{+}} \nabla' (z-z')  
\]
therefore
\[
\nabla \left[ \lim_{z'\rightarrow 0^{+} } \frac{z-z'}{|\boldsymbol{r}-\boldsymbol{r}'|^3} \right] = \lim_{z'\rightarrow 0^{+}} (z-z') \left[- \lim_{z'\rightarrow 0^{+} } \nabla'  \frac{1}{|\boldsymbol{r}-\boldsymbol{r}'|^3}\right] + \left[- \lim_{z'\rightarrow 0^{+}} \nabla' (z-z')\right] \lim_{z'\rightarrow 0^{+} } \frac{1}{|\boldsymbol{r}-\boldsymbol{r}'|^3}
\]
this is
\[
\nabla \left[ \lim_{z'\rightarrow 0^{+} } \frac{z-z'}{|\boldsymbol{r}-\boldsymbol{r}'|^3} \right] = -\lim_{z'\rightarrow 0^{+}} (z-z') \left[ \nabla' \left( \frac{1}{|\boldsymbol{r}-\boldsymbol{r}'|^3}\right) +   \frac{\nabla' (z-z')}{|\boldsymbol{r}-\boldsymbol{r}'|^3} \right]
\]
using again the product rule
\begin{equation}
\nabla \left[ \lim_{z'\rightarrow 0^{+} } \frac{z-z'}{|\boldsymbol{r}-\boldsymbol{r}'|^3} \right] = - \left[ \lim_{z'\rightarrow 0^{+}} \nabla' \frac{z-z'}{|\boldsymbol{r}-\boldsymbol{r}'|^3} \right].
\label{firstPropertyEq}
\end{equation}

\subsection{Second identity}
\label{AppendixSecondIdentiySubsection}
Here we shall demonstrate that
\[
\frac{\partial}{\partial z}\left[ \lim_{z' \to 0^{+}}\left(\frac{z-z'}{|\boldsymbol{r}-\boldsymbol{r}'|^3}\right) \right] = \frac{\partial}{\partial x'}\left[ \lim_{z' \to 0^{+}} \left(\frac{x-x'}{|\boldsymbol{r}-\boldsymbol{r}'|^3}\right) \right] + \frac{\partial}{\partial y'}\left[ \lim_{z' \to 0^{+}} \left(\frac{y-y'}{|\boldsymbol{r}-\boldsymbol{r}'|^3}\right) \right]
\]
Let us define
\begin{equation}
B_1(\boldsymbol{r},\boldsymbol{r}') := \frac{\partial}{\partial x'}\left[ \lim_{z' \to 0^{+}} \left(\frac{x-x'}{|\boldsymbol{r}-\boldsymbol{r}'|^3}\right) \right]\hspace{0.5cm}\mbox{and}\hspace{0.5cm}B_2(\boldsymbol{r},\boldsymbol{r}') := \frac{\partial}{\partial y'}\left[ \lim_{z' \to 0^{+}} \left(\frac{y-y'}{|\boldsymbol{r}-\boldsymbol{r}'|^3}\right) \right]
\label{definitionsB1andB2Eq}
\end{equation}
then
\[
B_1(\boldsymbol{r},\boldsymbol{r}') = \left[ \frac{\partial}{\partial x'}\lim_{z' \to 0^{+}} \left(x-x'\right) \right] \lim_{z' \to 0^{+}} \frac{1}{|\boldsymbol{r}-\boldsymbol{r}'|^3} +  \lim_{z' \to 0^{+}} \left(x-x'\right)  \frac{\partial}{\partial x'} \lim_{z' \to 0^{+}} \frac{1}{|\boldsymbol{r}-\boldsymbol{r}'|^5} 
\]
this is
\[
B_1(\boldsymbol{r},\boldsymbol{r}') = \lim_{z' \to 0^{+}}\left[- \frac{1}{|\boldsymbol{r}-\boldsymbol{r}'|^3} + \frac{3(x-x')^2}{|\boldsymbol{r}-\boldsymbol{r}'|^5}\right].
\]
Similarly
\[
B_2(\boldsymbol{r},\boldsymbol{r}') = \lim_{z' \to 0^{+}}\left[- \frac{1}{|\boldsymbol{r}-\boldsymbol{r}'|^3} + \frac{3(y-y')^2}{|\boldsymbol{r}-\boldsymbol{r}'|^5}\right].
\]
Hence
\[
B_1(\boldsymbol{r},\boldsymbol{r}') + B_2(\boldsymbol{r},\boldsymbol{r}') = \lim_{z' \to 0^{+}}\left[- \frac{2}{|\boldsymbol{r}-\boldsymbol{r}'|^3} + 3\frac{(x-x')^2 + (y-y')^2}{|\boldsymbol{r}-\boldsymbol{r}'|^5}\right] 
\]
Since
\[
(x-x')^2 + (y-y')^2 = \lim_{z' \to 0^{+}} \left[ |\boldsymbol{r}-\boldsymbol{r}'|^2-(z-z')^2 \right]
\]
then
\[
B_1(\boldsymbol{r},\boldsymbol{r}') + B_2(\boldsymbol{r},\boldsymbol{r}') = \lim_{z' \to 0^{+}}\left[- \frac{2}{|\boldsymbol{r}-\boldsymbol{r}'|^3} + 3\frac{|\boldsymbol{r}-\boldsymbol{r}'|^2-(z-z')^2}{|\boldsymbol{r}-\boldsymbol{r}'|^5}\right] = \lim_{z' \to 0^{+}}\left[ \frac{1}{|\boldsymbol{r}-\boldsymbol{r}'|^3} - 3\frac{(z-z')^2}{|\boldsymbol{r}-\boldsymbol{r}'|^5}\right] 
\]
This can be written as follows
\[
B_1(\boldsymbol{r},\boldsymbol{r}') + B_2(\boldsymbol{r},\boldsymbol{r}') = \lim_{z' \to 0^{+}} \left[\frac{\partial}{\partial z}(z-z')\right] \lim_{z' \to 0^{+}}\frac{1}{|\boldsymbol{r}-\boldsymbol{r}'|^3} + \lim_{z' \to 0^{+}}(z-z') \lim_{z' \to 0^{+}}\frac{-3(z-z')}{|\boldsymbol{r}-\boldsymbol{r}'|^5} 
\]
Now
\[
\frac{\partial}{\partial z} \lim_{z' \to 0^{+}} \frac{1}{|\boldsymbol{r}-\boldsymbol{r}'|^3} = \frac{\partial}{\partial z} \frac{1}{[(x-x')^2+(y-y')^2+z^2]^{3/2}} = \frac{-3 z}{|\boldsymbol{r}-\boldsymbol{r}'|^5} = \lim_{z' \to 0^{+}}\frac{-3 (z-z')}{|\boldsymbol{r}-\boldsymbol{r}'|^5}
\]
then
\[
B_1(\boldsymbol{r},\boldsymbol{r}') + B_2(\boldsymbol{r},\boldsymbol{r}') = \lim_{z' \to 0^{+}} \left[\frac{\partial}{\partial z}(z-z')\right] \lim_{z' \to 0^{+}}\frac{1}{|\boldsymbol{r}-\boldsymbol{r}'|^3} + \lim_{z' \to 0^{+}}(z-z') \frac{\partial}{\partial z} \lim_{z' \to 0^{+}} \frac{1}{|\boldsymbol{r}-\boldsymbol{r}'|^3} 
\]
using the product rule 
\[
B_1(\boldsymbol{r},\boldsymbol{r}') + B_2(\boldsymbol{r},\boldsymbol{r}') = \frac{\partial}{\partial z} \left[\lim_{z' \to 0^{+}} (z-z') \lim_{z' \to 0^{+}}\frac{1}{|\boldsymbol{r}-\boldsymbol{r}'|^3} \right] 
\]
replacing the definitions given by Eq.~(\ref{definitionsB1andB2Eq}) 
\[
\frac{\partial}{\partial x'}\left[ \lim_{z' \to 0^{+}} \left(\frac{x-x'}{|\boldsymbol{r}-\boldsymbol{r}'|^3}\right) \right] + \frac{\partial}{\partial y'}\left[ \lim_{z' \to 0^{+}} \left(\frac{y-y'}{|\boldsymbol{r}-\boldsymbol{r}'|^3}\right) \right] = \frac{\partial}{\partial z} \left[\lim_{z' \to 0^{+}} (z-z') \frac{1}{|\boldsymbol{r}-\boldsymbol{r}'|^3} \right].
\]

\end{appendices}

%\begin{thebibliography}{99} % Bibliography - this is %intentionally simple in this template

%\bibitem{andreotti1997studying} B. Andreotti, \emph{Studying Burgers' models to investigate the physical meaning of the alignments statistically observed in turbulence}, Phys. Fluids \textbf{9} : 3, March (1997)

%\bibitem{cohl1999compact} Cohl, Howard S., and Joel E. Tohline, \emph{A compact cylindrical Green's function expansion for the solution of potential problems}, The astrophysical journal \textbf{527} : 86 - 101 (1999) %DOI: https://doi.org/10.1086/308062

%\bibitem{abramowitz1965handbook} Abramowitz, Milton, and Irene A. Stegun. \emph{Handbook of Mathematical Functions With Formulas, Graphs, and Mathematical Tables.} (1964).

%\end{thebibliography}

%croos section karlie
%GOOD: http://www.eumetrain.org/satmanu/CMs/TrCyAt/print.htm 
%https://physics.stackexchange.com/questions/275799/why-is-the-eye-of-a-cyclone-a-forced-vortex
%http://www.chanthaburi.buu.ac.th/~wirote/met/tropical/textbook_2nd_edition/navmenu.php_tab_9_page_7.1.0.htm
%http://www.atmos.umd.edu/~dalin/andrew/part2.html
%https://nptel.ac.in/courses/119102007/2
%http://www.911omissionreport.com/steering_hurricanes.html
%https://www.youtube.com/watch?v=_brY_9ME8iE brooks
\end{document}